\documentclass[aps,prx,superscriptaddress,twocolumn,longbibliography]{revtex4-2}

\usepackage{appendix}

\usepackage[T1]{fontenc}

\usepackage[pdftex]{graphicx}
\usepackage{color}
\usepackage{amsmath}
\usepackage{subcaption}

\usepackage{setspace}

\usepackage{amsmath}
\usepackage{xfrac}
\usepackage{xcolor}

\usepackage{dcolumn}   
\usepackage{amsfonts}
\usepackage{booktabs}
\usepackage{siunitx}
\usepackage{multirow}
\usepackage{mathtools}
\usepackage{multirow}

\usepackage{lineno,hyperref}

\newcommand{\MARS}{Aix-Marseille Universit\'{e}, CNRS, CINaM UMR 7325, Campus de Luminy, 13288 Marseille, France}

\newcommand{\QUEENS}{Department of Mechanical and Materials Engineering, Queen's University, Kingston, Ontario, Canada}

\begin{document}

	\title{Interstitialcy-based reordering kinetics of Ni$_3$Al precipitates in irradiated Ni-based super alloys}
	
	\author{Keyvan Ferasat}
	\email{17kf7@queensu.ca}
	\affiliation{\QUEENS}
	
	\author{Thomas D. Swinburne}
	\email{swinburne@cinam.univ-mrs.fr}
	\affiliation{\MARS}
	
	\author{Peyman Saidi}
	\email{peyman.saidi@queensu.ca}
	\affiliation{\QUEENS}
	
	\author{Mark R. Daymond}
	\email{mark.daymond@queensu.ca}
	\affiliation{\QUEENS}
	
	\author{Zhongwen Yao}
	\email{yaoz@queensu.ca}
	\affiliation{\QUEENS}
	
	\author{Laurent Karim B\'{e}land}
	\email{laurent.beland@queensu.ca}
	\thanks{Corresponding author.}
	\affiliation{\QUEENS}

	\date{\today}
	
	\begin{abstract}
		
		Neutron irradiation tends to promote disorder in ordered alloys through the action of the thermal spikes that it generates, while simultaneously introducing point defects and defect clusters. As they migrate, these point defects will promote reordering of the alloys, acting against irradiation-induced disordering. In this study, classical molecular dynamics
		and a highly parallel accelerated sampling method are used to study the reordering kinetics of Ni$_3$Al under the diffusion of self-interstitial atoms (SIA). By monitoring the order parameter and potential energy from atomistic simulations, we show that the SIA acts as a reordering agent in Ni$_3$Al. A mean-field rate theory model of the interstitialcy-based reordering kinetics is introduced, which reproduces simulation data and predicts reordering at temperatures as low as 500 K.
		
	\end{abstract}
	
	\maketitle
	
	\section{Introduction}
	
	Metallic alloy systems characterized by large heat of mixing tend to form ordered compounds. The introduction of such ordered secondary-phase particles is a common strategy employed to strengthen alloys. The Ni$_3$Al ($\gamma^{'}$) phase in Ni-based superalloys is one such example. It has a considerable negative mixing enthalpy that extends the ordered phase up to its melting temperature \cite{cahn2013physics, porter2009phase, ewert1998ion}. The intermetallic compound Ni$_3$Al not only enhances high temperature strength, but also improves the oxidation resistance of Ni-based superalloys. Therefore, Ni-based alloys such as X-750 and 718 are used for in-core and out-core components in the pressurized water reactor (PWR), boiling water reactor (BWR), and Canada deuterium uranium reactor (CANDU). In-core Ni-based parts are subjected to high neutron exposure: the X-750 spacers in CANDU reactors need to withstand neutron environment for periods of decades \cite{davis2000special, griffiths2014review_Ni}. Neutron irradiation can cause disordering of ordered phases and affect mechanical properties, similarly to other processes that introduce external energy to a system, such as severe plastic deformation and rapid solidification \cite{yavari1989rapid, yavari1993reordering, scherrer1998nmr, dimitrov1994evolution, ewert1998ion, pochet1995order}. In addition to disordering Ni$_3$Al ($\gamma^{'}$) phase in X-750 Ni-based superalloy \cite{griffiths2014review_Ni, zhang2014stability}, neutron--and ion--irradiation introduces point defects and defect clusters to the materials. Because of the strong thermodynamic driving force towards order, point defects will act as reordering agents as they diffuse \cite{bourdeau1994disordering, zhang2013microstructural, zhao2019diffusion}.
	
	Molecular dynamic (MD) collision cascade studies indicate that disordering is a consequence of cascade-induced thermal spikes \cite{spaczer1994computer, spaczer1995evidence, gao2000temperature, ye2010atomistic, zhang2014radiation, lee2015atomistic}. A number of experimental studies verified by rate theory and on-the-fly kinetic Monte Carlo simulations indicate that mono-vacancies act as a reordering agent at temperatures more than 750 K \cite{oramus2003ordering, martinez2016atomistic}. Recently, molecular dynamics simulations in ordered and disordered (NiCo)$_3$Al and Ni$_3$Al-Ni interface effect were reported \cite{zhao2021structural, zhao2020atomistic}. Vacancy and self-interstitial atom (SIA) diffusion exhibit different behavior in ordered and disordered structures: vacancies migrate faster in a disordered structure while SIAs are faster in ordered ones. Moreover, transmission electron microscopy (TEM) of irradiated samples with and without helium implantation coupled to atomistic simulations and a rate theory model showed that helium can act as a reordering agent at lower temperatures \cite{saidi2020effect}. There are also reports that reordering occurs at temperatures less than 750 K even without helium \cite{liu1983irradiation, bourdeau1994disordering, zhang2014stability, zhang2020effects}. Scientific interest in reordering kinetics led to the development of various mean-field models for order-disorder reactions in Ni$_3$Al phase \cite{abromeit1996reorder, njah1990kinetics, njah1999modelling, ewert2000reordering}. Reordering by vacancies is well studied and understood, and applicable models are presented in Refs. \cite{bourdeau1994disordering, ewert2000reordering, njah1990kinetics}. Moreover, most of these studies tried to enlighten the role of SIAs at temperatures lower than 750 K in which vacancy diffusion is not effective. These reports indicate that SIAs are an efficient reordering agent when the annihilation process is dominated by rapid mutual recombination \cite{njah1999modelling}. Although all of these studies consider a similar irradiation disordering rate, the ordering rate varies due to differences in the thermodynamic driving forces. The motivation of this study is the lack of a good reordering rate model for SIAs as well as their effectiveness as a reordering agent in systems disordered by irradiation.
	
	In this contribution, we use brute force MD and the massively parallel accelerated sampling method \texttt{TAMMBER} \cite{swinburne2018self,swinburne2020automated} to characterize the energy landscape of SIA diffusion in disordered Ni$_3$Al and elucidate the role of SIAs in reordering. Furthermore, a model that reproduces our simulation results is introduced to predict the ordering rate for SIAs in the Ni$_3$Al system.	
		
	\section{Methods}
	
	\subsection{Molecular Dynamics}
	
	We introduced a SIA in Ni$_3$Al system with different order parameters, and tracked its trajectory in the isothermal–isobaric ensemble using MD simulations. The \texttt{LAMMPS} \cite{plimpton1995fast} software package was used to perform the MD simulations. The interatomic potential used in this study is the same as Ref. \cite{saidi2020effect} which is developed based on Refs. \cite{mishin2004atomistic, skirlo2012role, torres2017atomistic}. We chose this potential \cite{mishin2004atomistic} because it is focused and narrowed for Ni$_3$Al while the newer potential \cite{purja2009development} is more universal for Ni-Al systems. SIA transport is studied for a combination of different initial long range order (LRO) parameters at six different temperatures (650, 700, 750, 900, 1000 \& 1100 K). Our simulation box contains 3000 Ni and 1000 Al atoms. We first arrange them in a L1$_2$ crystal structure--a fully ordered structure. Then Ni and Al atoms are shuffled to reach the desired LRO parameters from 0\% to a fully ordered crystal. Based on the Bragg-Williams LRO parameter \cite{hoyt2011phase}, either of equations $\xi =  \frac{\text{r}_{\text{Al}}-\text{C}_{\text{Al}}}{1-\text{C}_{\text{Al}}} = \frac{\text{r}_{\text{Al}}-0.25}{1-0.25}$ \& $\xi =  \frac{\text{r}_{\text{Ni}}-\text{C}_{\text{Ni}}}{1-\text{C}_{\text{Ni}}} = \frac{\text{r}_{\text{Ni}}-0.75}{1-0.75}$ can be used to make disorder structures from an ordered one, where $\xi$ is the LRO parameter and r$_{\text{Al}}$ is the probability of Al atoms occupying the Al sublattice. In this study, equation $\xi = \frac{\text{r}_{\text{Al}}-0.25}{1-0.25}$ and "type/subset" command in \texttt{LAMMPS} \cite{plimpton1995fast} is used to shuffle exactly the required number of Al and Ni atoms on each basis of fcc structure. Finally, a Ni atom is added to form a dumbbell SIA, i.e. having in total 4001 atoms in the box.
	
	At low (650, 700 \& 750 K) and high (900, 1000 \& 1100 K) temperatures, 200 and 10 ns MD simulations are respectively done to obtain enough statistics for mean square displacement (MSD). In addition to MSD, the change in potentials energy and short range order (SRO) parameter are recorded to evaluate the role of SIA in reordering. SRO parameters are calculated by analyzing the \texttt{LAMMPS} dump files using \cite{porter2009phase}:
	
	\begin{equation}\label{SRO}
		\text{SRO} =  \frac{\text{P}_{\text{Ni-Al}}-\text{P}^{\text{rand}}_{\text{Ni-Al}}}{\text{P}^{\text{max}}_{\text{Ni-Al}}-\text{P}^{\text{rand}}_{\text{Ni-Al}}},
	\end{equation}
	
	where $\text{P}_{\text{Ni-Al}}$ is the number of Ni-Al bonds. The number of Ni-Al bonds are counted by \textit{ovito} \cite{stukowski2009visualization} with a fixed cutoff of 3.1\AA. $\text{P}^{\text{max}}_{\text{Ni-Al}}$ is the maximum number of Ni-Al bonds for a fully ordered structure (12000 for a system with 4000 atoms), and $\text{P}^{\text{rand}}_{\text{Ni-Al}}$ is the number of Ni-Al bonds in a fully random disordered structure. In addition to equation \ref{SRO}, one can calculate SRO as explained in Ref. \cite{cowley1950approximate}. $\text{P}^{\text{rand}}_{\text{Ni-Al}}$ is equal to 9000 for a system with 4000 atoms as calculated based on Ref. \cite{cowley1950approximate}. SRO calculations and its relationship with LRO are explained in the supplementary materials. 
	
	\subsection{\texttt{\texttt{TAMMBER}} Simulations}
	At temperatures lower than 650 K, SIA diffusion cannot be captured by standard MD due to timescale limitations, requiring the use of accelerated methods. We used the \texttt{\texttt{TAMMBER}} code to study the energy landscape of SIA diffusion in three different disordered Ni$_3$Al configurations (40, 60 \& 80 LRO). \texttt{\texttt{TAMMBER}} \cite{swinburne2018self, swinburne2020automated} is a massively parallel accelerated sampling scheme which distributes hundreds to thousands of MD `workers' simultaneously to rapidly discover networks of local energy minima and connecting minimum energy paths, utilizing the nudged elastic band \cite{Henkelman2000NEB, Henkelman2000-improved-NEB} and transition state theory to construct the transition catalog. 
	As any finite sampling effort is unlikely to discover all available minima and connections, \texttt{TAMMBER} uses Bayesian methods to estimate the `unknown' rate for each state, being the difference between the discovered escape rate and the true escape rate at a given temperature. This information is then used to autonomously assign where workers start MD simulations, such that the predicted akMC model quality increases as fast as possible \cite{swinburne2018self}. This facilitates parallel application as the optimal sampling tasks (based on the available sampling data) are determined with no end user supervision. Group theoretical techniques have been used to massively accelerate and automate the calculation of defect diffusivities in ordered alloys \cite{swinburne2020automated}. In the present work, \texttt{TAMMBER} was used to rapidly characterize the energy landscape of SIA diffusion in disordered Ni$_3$Al, through the discovery of many thousands of distinct SIA migration pathways. The activation barriers were used to parametrize a mean-field analytical model, which we describe in section \ref{AnalyticalModel}.
	
	\section{Results}
	
	\subsection{Molecular Dynamics}
	
\begin{figure*}[!ht]
		\centering
		\includegraphics[scale=0.5]{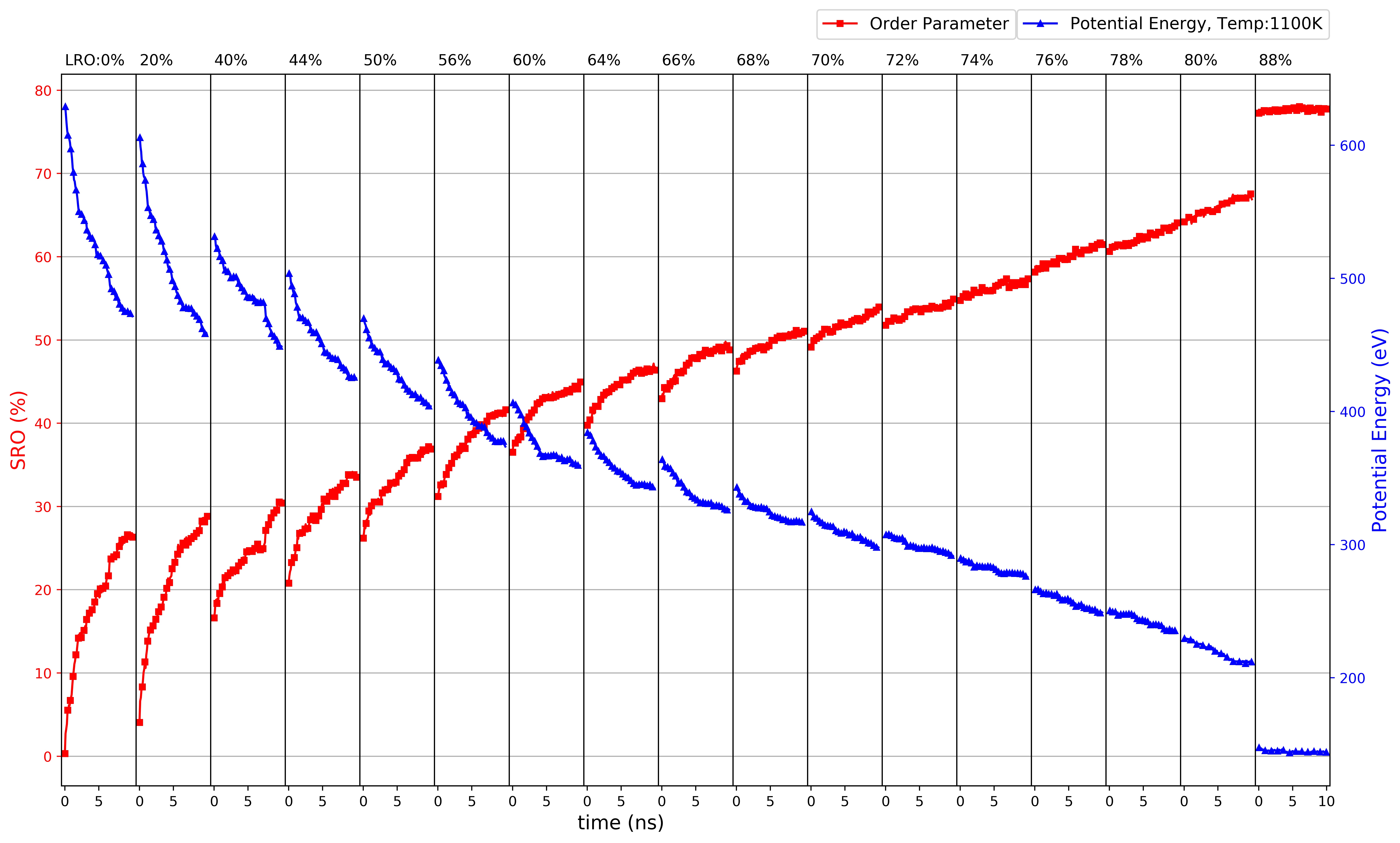}
		\caption{Change in the potential energy and SRO parameter as a function of time at 1100 K. Each of the small segments are 10 ns MD simulations (the MD segments ran at temperatures cooler than 900K were 200 ns each). The LRO parameter of the initial configurations are indicated on the top of each segment.}
		\label{pot_ord}
	\end{figure*}

	Fig. \ref{pot_ord} illustrates the change in SRO parameter and potential energy as a function of time at 1100 K given different initial configurations, i.e. different LRO parameters. In each run, the potential energy decreases and the SRO parameter increases as the SIA migrates. These changes are due to formation of new more favorable Ni-Al bonds. Analogous plots at other temperatures (650, 700, 750, 900 \& 1000 K) are shown in the supplementary materials. In general, the rate of reordering decreases as the order parameter increases.
	
\begin{figure*}[bht]
		\centering
		\begin{subfigure}[b]{.4\linewidth}
			\includegraphics[scale=0.5]{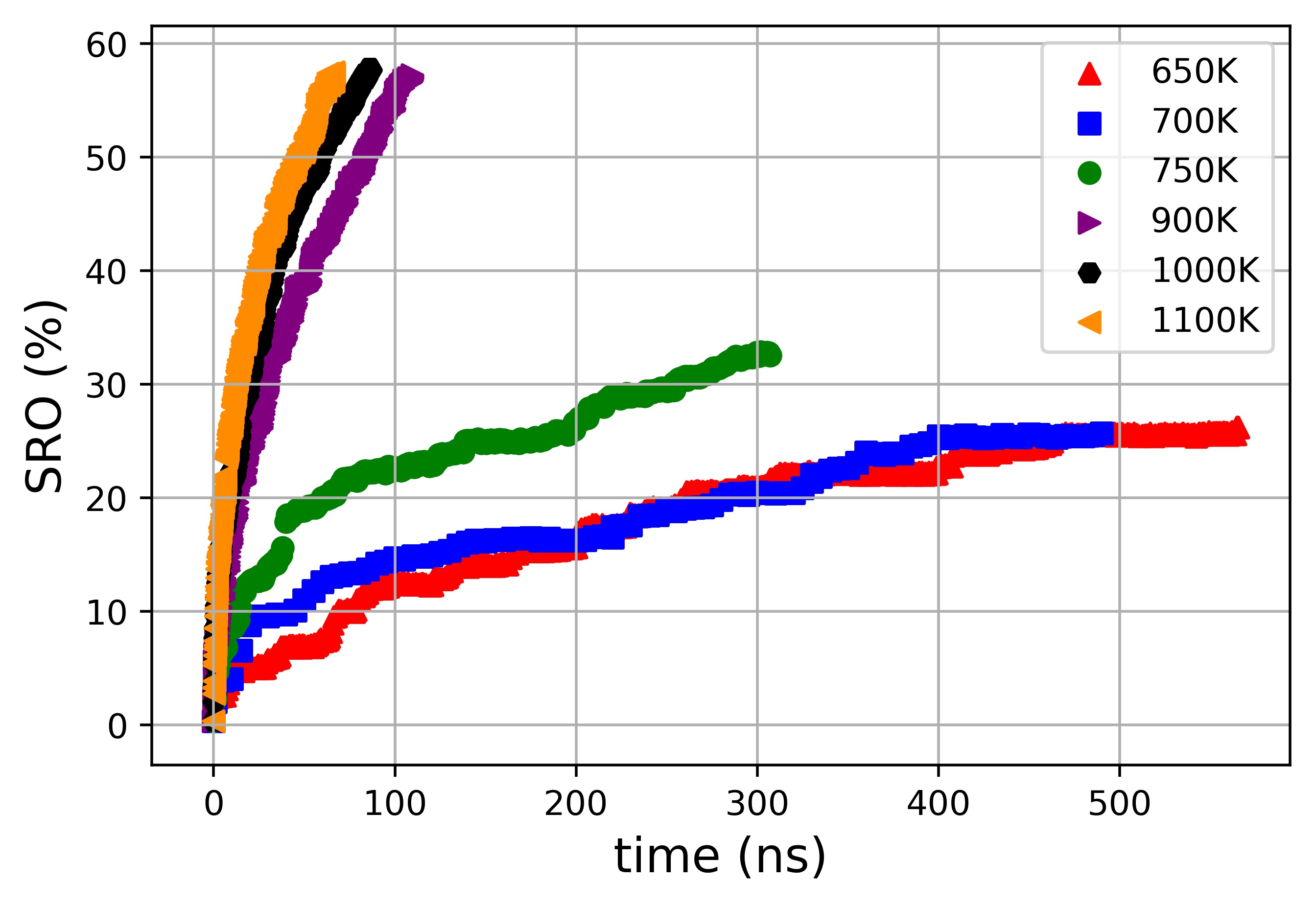}
			\caption{SRO as a function time}
		\end{subfigure}
		\begin{subfigure}[b]{.4\linewidth}
			\includegraphics[scale=0.5]{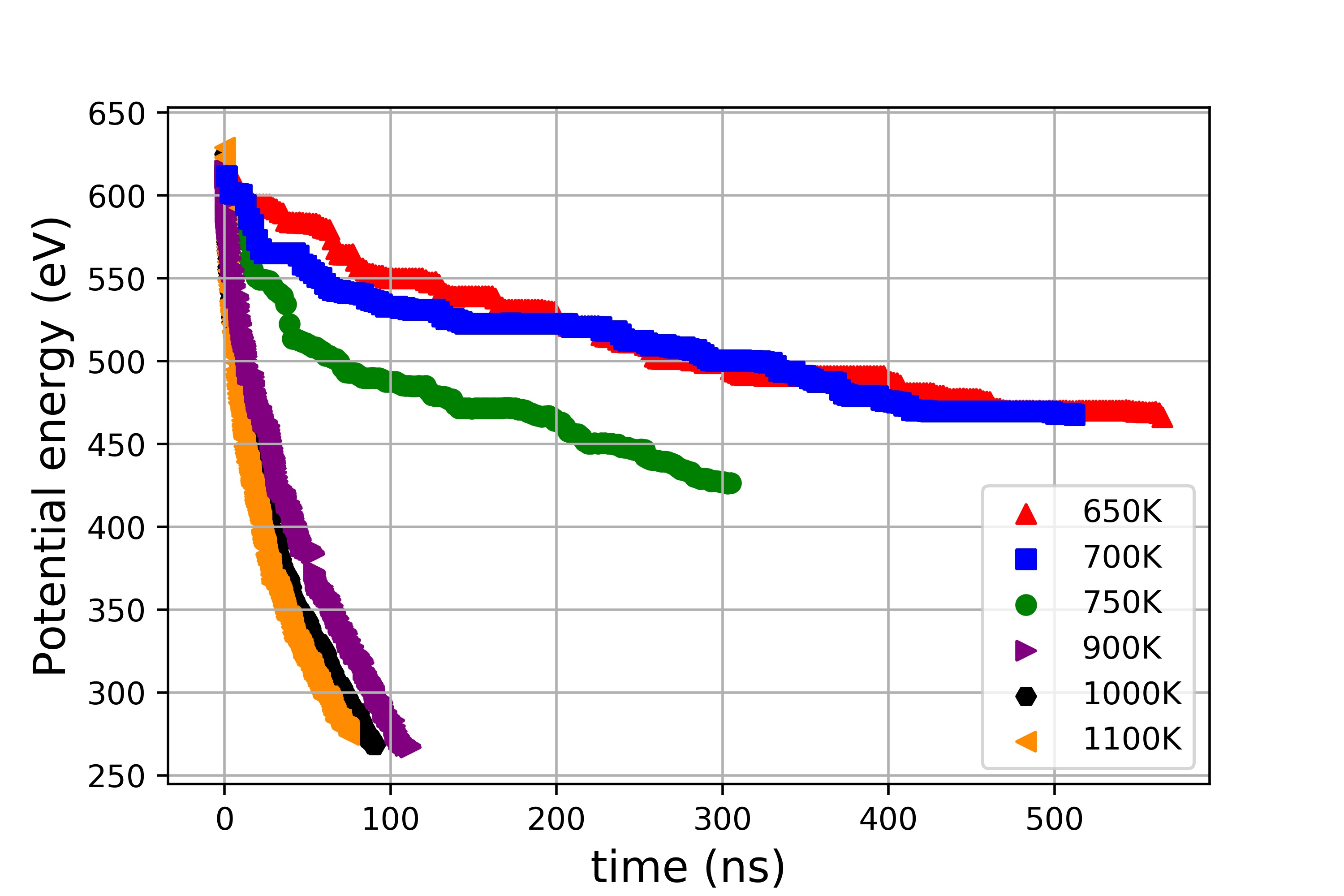}
			\caption{Potential energy as a function of time}
		\end{subfigure}
		\caption{SRO and potential energy after splicing MD segments that overlapped in SRO and potential energy. At 900, 1000 \& 1100 K, we spliced trajectories up to 74\% LRO. At 650 \& 750 K they were spliced up to 44\% LRO, and at 700K, we spliced them up to 40\% LRO.}
		\label{SRO-PE}
	\end{figure*}

	As can be seen in Fig. \ref{pot_ord}, the observed SRO parameter and potential energy of different MD runs overlap. Based on this observation, we spliced different short MD segments--note that these "short" MD segments can be as long as 200 ns each--to generate longer SRO parameter vs time plots. This lets us construct trajectories that describe reordering for up to 550 ns. This scheme is applied to all the temperatures considered in this study; the corresponding graphs are shown in Fig. \ref{SRO-PE}.
	
	Diffusion coefficients were calculated by linear regression of atomic square displacement (SD) vs. time. Particularly at higher temperatures, SD plots are fairly linear and contain enough statistics to calculate diffusion coefficient with small error bars. Error bars were calculated by trajectory decomposition. SD as a function of time at 1100 K is shown in Fig. \ref{square} and for other temperatures are indicated in supplementary materials. 
	
	D$^*$--the total atomic diffusion coefficient--as a function of initial LRO at 1100 K is shown in Fig. \ref{diff_active}(a) and at other temperatures are shown in supplementary materials. Moreover, to check the effect of initial configurations with the same LRO (but slightly different SRO) on the calculated D$^*$, 22 different initial configurations for each LRO parameter (0\% to 80\%) at 1100 K are simulated for 10 ns. The D$^*$ distribution for these 22 different initial conditions are reported in Fig. \ref{diff_active}(a). These calculations indicate that the calculated D$^*$ is essentially independent of exact initial atomic configuration chosen, given a certain LRO parameter. Interestingly, the variance induced by using different initial configurations is of the same order of magnitude as the variance calculated using trajectory decomposition.
	
	As can be seen in Figs. \ref{square} \& \ref{diff_active} (a), D$^*$ and D$_{Ni}^*$ (Ni tracer) are roughly independent of the degree of order given LRO parameters ranging from 0\% to 70\%. From 70\% to 100\% LRO parameters, D$^*$ and D$_{Ni}^*$ are roughly doubled.  As for  D$_{Al}^*$ (Al tracer), it decreases markedly as the order parameter increases. In systems with LRO parameters larger than 80 \%, D$_{Al}^*$ is negligible.
	
	\begin{figure*}[!ht]
		\centering
		\includegraphics[scale=0.6]{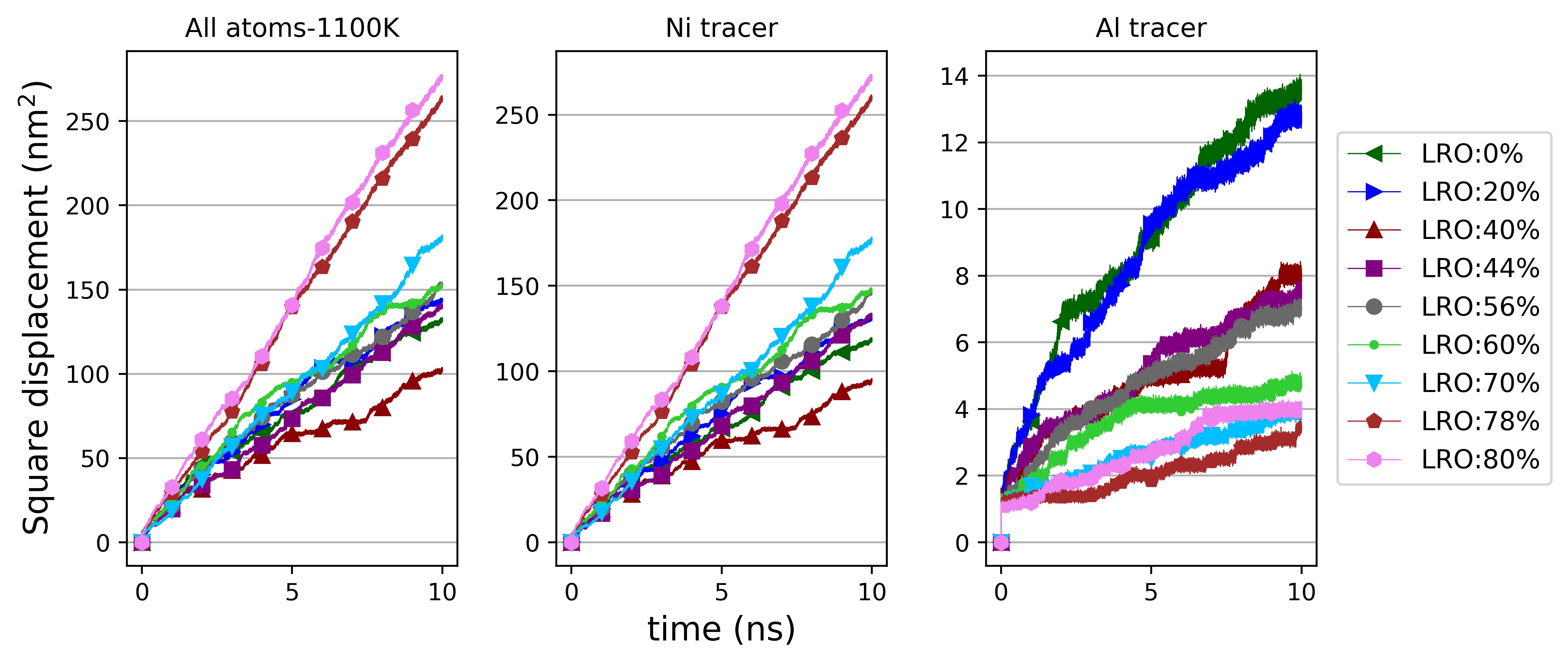}
		\caption{Atomic square displacement as a function of time at 1100 K, given different initial configurations (LRO parameters).}
		\label{square}
	\end{figure*}
	
	Effective migration energy (Q$_{eff}$) as a function of LRO parameter is obtained by an Arrhenius treatment of tracer diffusion coefficients at different temperatures, and is shown in Fig. \ref{diff_active}(b). Q$_{eff}$ is roughly independent of LRO parameter between 0\% and 60 \% LRO parameter. It has a value of 0.6 eV. Note that this value is different from the value reported by Zhao \& Osetsky \cite{zhao2021structural} (0.32 eV), likely because they used a different interatomic potential. At higher order parameters, the effective migration energy decreases, and reaches 0.11 eV in the fully ordered system. 
	
	\begin{figure*}[bht]
		\centering
		\begin{subfigure}[b]{.4\linewidth}
			\includegraphics[scale=0.5]{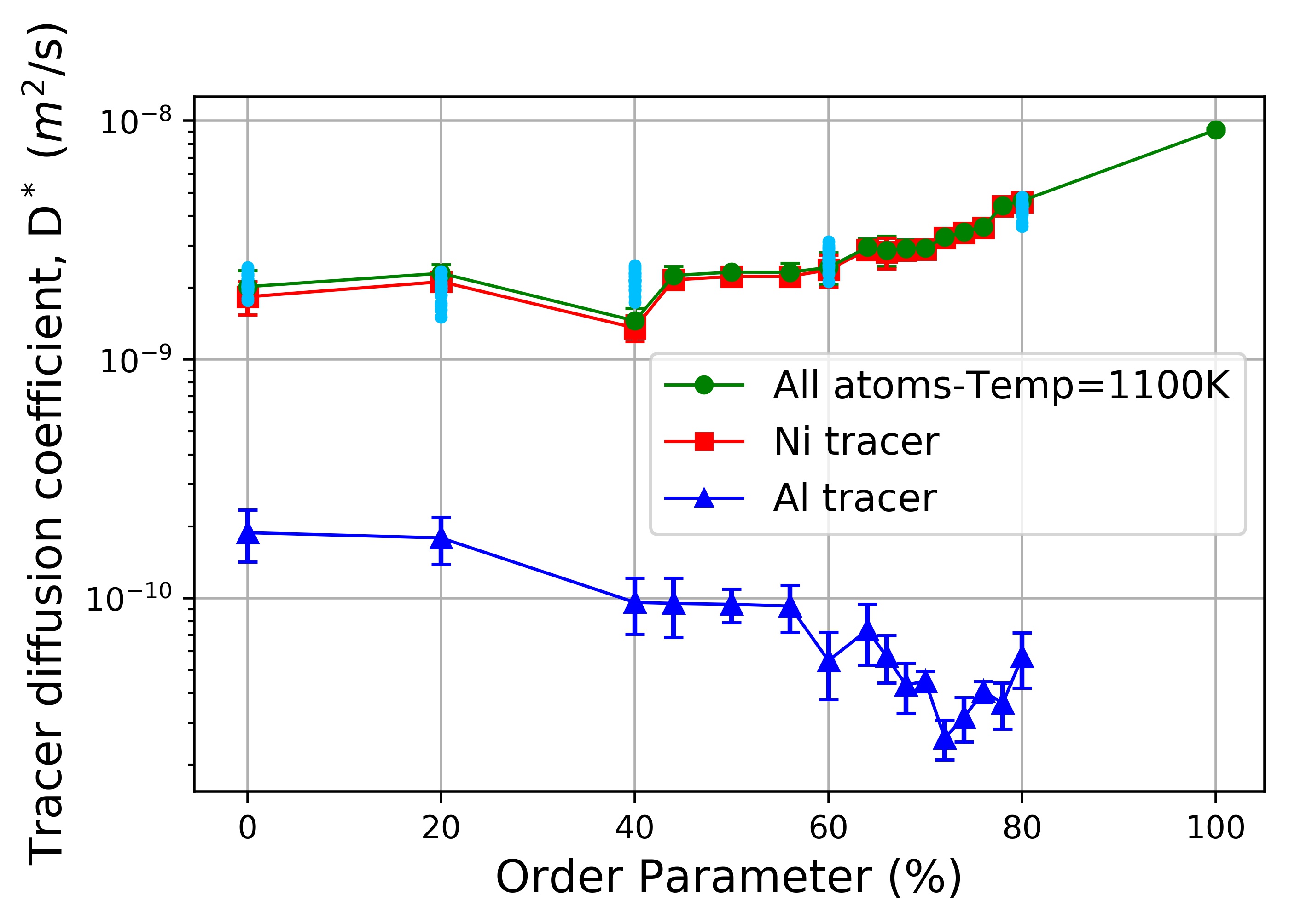}
			\caption{Total atomic diffusion coefficient at 1100 K.}
		\end{subfigure}
		\begin{subfigure}[b]{.4\linewidth}
			\includegraphics[scale=0.5]{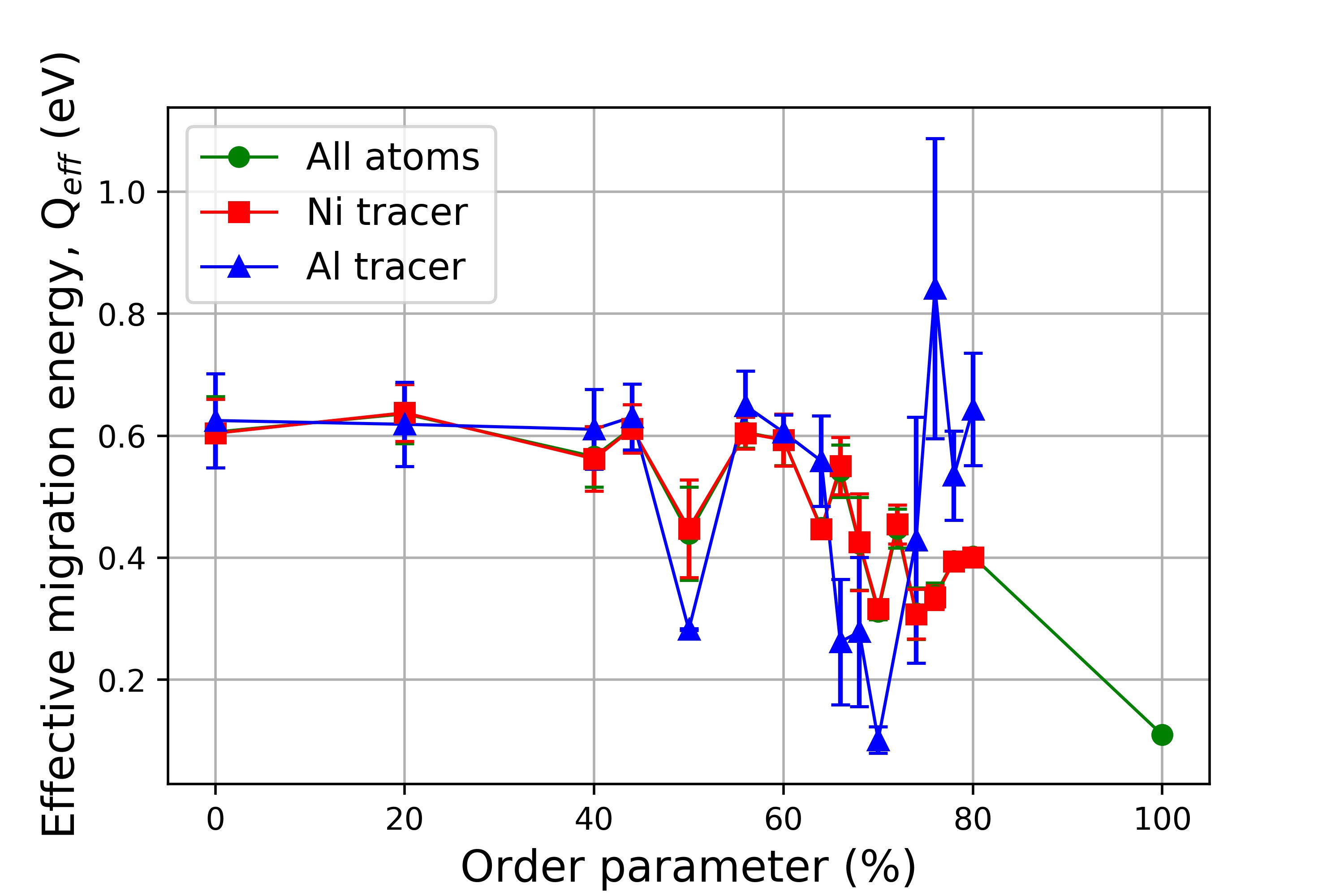}
			\caption{Effective migration energy}
		\end{subfigure}
		\caption{D$^*$ and Q$_{eff}$ as a function of order parameter show a flat/non-monotonic behavior. (a) In order to check the effect of initial configuration, 10ns MD simulations at 1100 K are performed given 22 different initial configurations. These D$^*$ are shown with light blue circles. The error bars are calculated by trajectory decomposition. (b) Effective migration energies, Q$_{eff}$, are calculated through an Arrhenius treatment of D$^*$.}
		\label{diff_active}
	\end{figure*}
	
	\subsection{\texttt{TAMMBER} simulations}
\begin{figure*}[!ht]
		\centering
		\includegraphics[width=\textwidth]{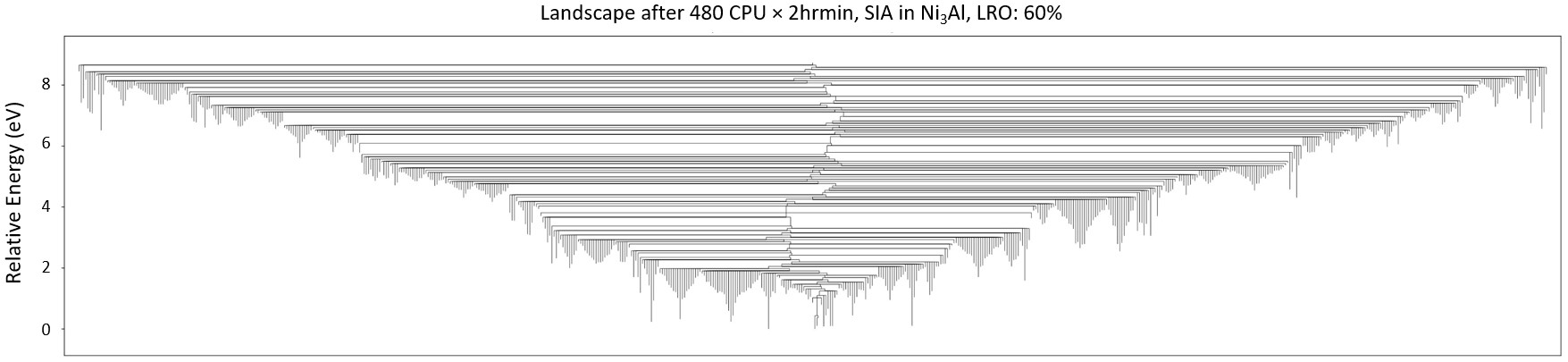}
		\caption{An illustration of the energy landscape explored by \texttt{TAMMBER}, in the form of a disconnectivity graph. This \texttt{TAMMBER} run was initiated starting from a configuration with an 60\% LRO parameter}
		\label{disconnect}
	\end{figure*}

	Given an initial disordered structure of 40\%, 60\% or 80\% LRO, \texttt{TAMMBER} was deployed for 2 hours on 480 cores, producing many thousands of SIA migration events and an akMC model valid for a range of temperatures, in particular low temperatures inaccessible to MD simulation timescales \cite{swinburne2018self}. This computational effort of approximately 1000CPUh per LRO is comparable to standard MD simulation for each LRO and temperature, each of which requires around one week of simulation time, due to limited scalability. Representative simulation results for 60\% LRO are illustrated in Fig. \ref{disconnect} as a disconnectivity graph \cite{wales2003energy}, where each vertical line represents a local minima, whilst the tree structure displays the lowest energy saddle point between any two minima. The disconnectivity graph reveals a rich and complex energy landscape despite the superficial simplicity of the Ni$_3$Al system, reminiscent of $\alpha$ and $\beta$ relaxations typically observed in metallic glasses \cite{lan2017hidden}.  
	
\begin{figure*}[!ht]
		\centering
		
		\includegraphics[width=0.49\textwidth]{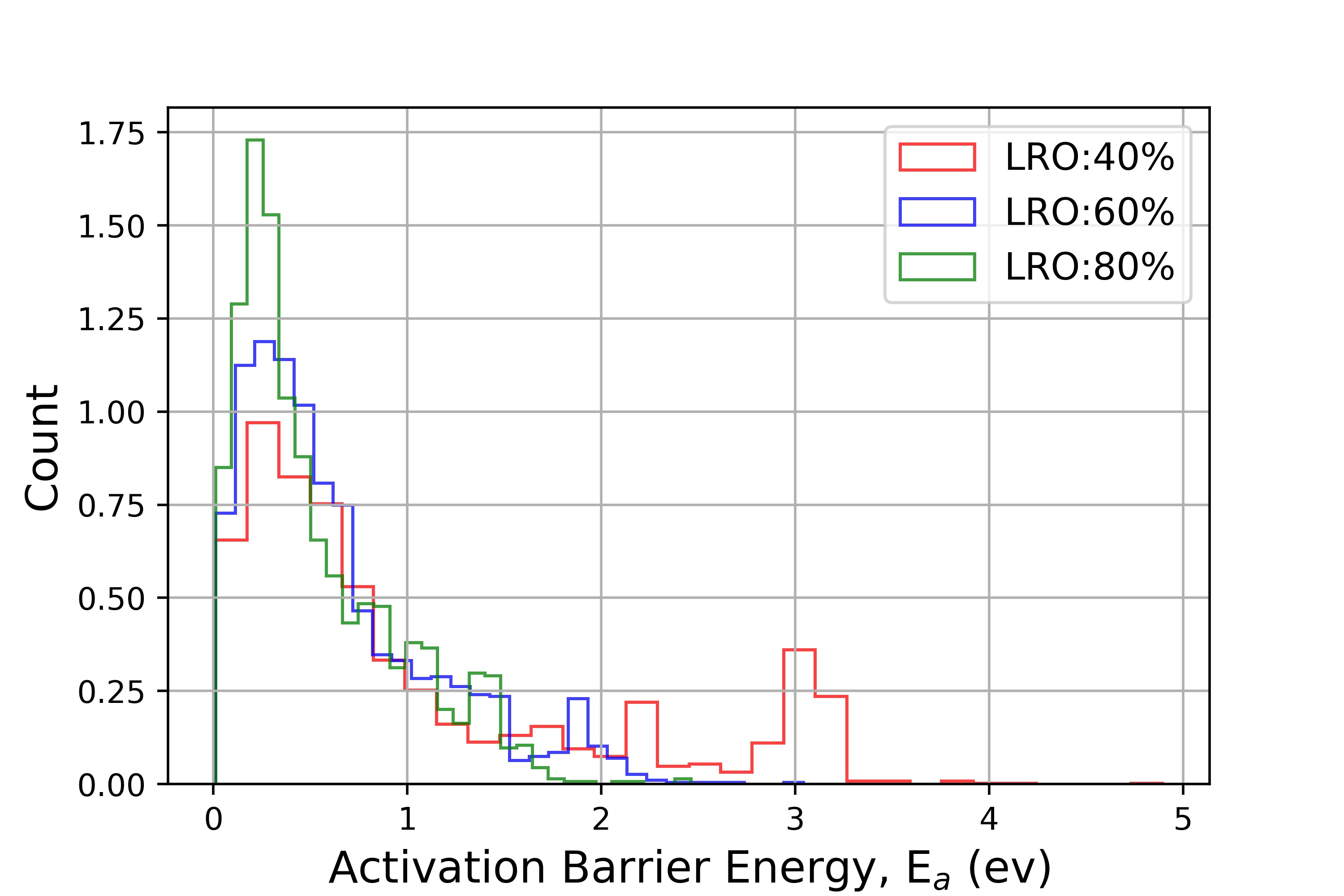}
		\includegraphics[width=0.49\textwidth]{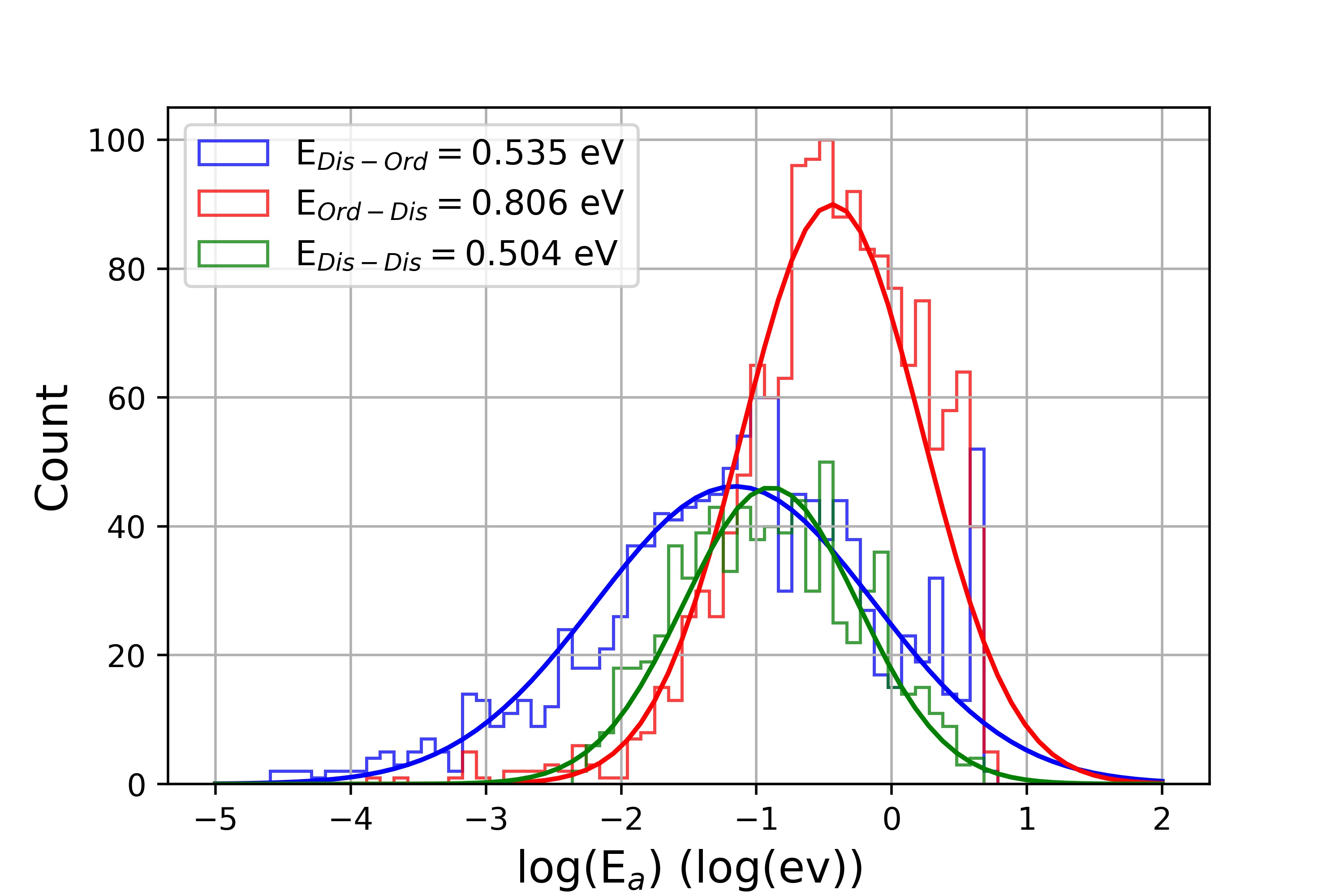}
		\caption{Activation energies from \texttt{TAMMBER} simulations. Left: The distribution of activation energies found by three \texttt{TAMMBER} runs, each exploring the energy landscape of configurations characterized by a different initial LRO parameter. Right: The distribution of the activation energies found by \texttt{TAMMBER} (combining 40\% \& 60\% LRO parameter runs). We illustrate the activation energy of transitions that increase SRO, that decrease SRO, and that do lead to no change in SRO. The mean value of these log-normal distributions are shown in legend.}
		\label{migrate}
	\end{figure*}

	A statistical analysis of the migration barrier database allows insight into the reordering kinetics. Based on the change in potential energy and SRO during a given migration event, migration barriers of SIA jumps can be separated into three categories: the jumps that increase SRO, the ones that decrease SRO, and finally the jumps that do not change SRO.
	Fig. \ref{migrate}(a) shows all of the migration barriers calculated with \texttt{TAMMBER}. In the next step, SIA jumps are separated in the mentioned three categories based on increasing, decreasing, and not changing SRO. 
	
	As SRO is a local metric,  we also collated all barriers obtained in the 40\% LRO and 60\% LRO runs to improve our regression analysis. Barriers more than 2 eV are ignored since they are not related to dumbbell rotation-transfer movement to the first-nearest-neighbor sites. 80\% data were not considered since the diffusion kinetics changes once the LRO parameter significantly surpasses 60\% (see Fig. \ref{diff_active} and the previous section's discussion). As indicated in Fig. \ref{migrate}(b), the migration barriers are well-described by a log-normal distribution. E$_{Dis-Ord}$, E$_{Ord-Dis}$, E$_{Dis-Dis}$ indicates the migration barriers when SRO increases, decreases, and does not change, respectively. Being in a disordered structure is the reason why "Dis-Dis" subscript is used when SRO does not change.
	
	\section{Analysis, Discussion \& Analytical model}
	
	The reordering rate decreases significantly as the LRO and SRO parameters increase. For instance, Fig. \ref{pot_ord} shows that the SRO parameter increased by 27 percentage points in 10 ns starting from a completely disorered system; it increased by less than 2 percentage points in 10 ns starting from a 88\% LRO paramater state. This indicates that in the latter more ordered state, the SIA mostly diffuses through a continuous Ni path, which does not affect SRO and potential energy. Reordering only takes place when Al atoms contribute to SIA jumps. A comparison between Figs. \ref{pot_ord} and \ref{diff_active}(b) indicates that while the effective SIA migration barrier is lower at higher order parameters, the reordering rate decreases. It is well-established that the Ni-Ni is the most stable dumbbell in the Ni$_3$Al system \cite{gao2000temperature, caro1990threshold}. Moreover, it can be geometrically shown that if the preferred chemical species of a SIA dumbbell in a binary fcc crystal has a concentration of more than 20-25\%, a continuous path exists for SIA diffusion \cite{xu2014simultaneous}. In Ni$_3$Al, 75\% Ni concentration guarantees the existence of a continuous migration path for Ni-Ni dumbbells. Hence, due to a lower driving force of reordering at higher LROs and existence of continuous Ni-channels, the contribution of Al atoms in SIA jumps are reduced and the reordering rate decreases. Similarly, this can be observed from Fig. \ref{square}(c), as Al atoms contribute the most to square displacement when the LRO parameter is less than 40\%.
	
	The fact that D$^*$ is independent of order parameter from 0\% to 70\% LRO parameter can be explained based on the driving force for diffusion and SIA dumbbell configurations. Localized diffusion traps are created when atoms are shuffled to generate the disordered structures. In other words, the continuous Ni-Ni path present in the fully ordered structure may be broken. It is well understood that such geometrical considerations are an important factor controlling diffusion kinetics \cite{mahmoud2018long}. In our study, two main factors control diffusion of SIAs. Disorder introduces a thermodynamic driving forces towards order, in conjunction with a higher chance of breaking continuous Ni-Ni paths. The competition between these two factors leads to D$^*$ and effective migration energies Q$_{eff}$ being independent of order parameter from 0\% to 70\% LRO parameter. In other words, when the LRO is low, there is a large driving force towards reordering, which promotes escaping from configurational traps. As the LRO increases from 0\% to 70\% LRO, the concentration of traps decreases, in tandem with the driving force for leaving them. This leads to D$^*$ being independent of LRO parameter in this regime. When the LRO parameter reaches 70\%, traps become very rare--which is consistent with percolation theory-- and D$^*$ increases as the LRO parameter increases.
	
	The fact that the total Q$_{eff}$ of SIA is 10-50\% of the vacancy-mechanism barrier \cite{gopal2012first} is not enough to guarantee that the interstitialcy mechanism will lead to ordering at low temperatures. As we discussed, migration of the SIA does not automatically introduce new Ni-Al bonds to the system. In the ordered and close-to-ordered (80\%+ LRO parameter) configurations, the dumbbell rarely increases the number of Ni-Al bonds as it moves. In other words, a small migration barrier is a necessary but not sufficient condition for SIAs to be effective reordering agents at low temperatures. 
	
	Our MD results suggest that SIAs could be a possible reordering agent at temperatures as low as 650K. However, to study even lower temperatures, such as 500 K, a predictive analytical model is needed. In the next subsection, such a model is introduced. 
	
	\subsection{SIA reordering model}
	\label{AnalyticalModel}
	Previous attempts to model vacancy- and interstitialcy-mediated reordering employ an efficiency term that largely controls reordering \cite{njah1999modelling, ewert2000reordering}. Following these previous models, here, the reordering rate is considered as follows:
	\begin{equation}\label{model}
		\frac{d\eta}{dt} = \alpha(\eta,T)\frac{dj}{dt},
	\end{equation}
	
	where $\eta$ is the SRO parameter, and $\frac{dj}{dt}$ is the jump rate. $\alpha$ is the jump efficiency, which is a function of both SRO and temperature.
	
	In order to the find jump efficiency, $\alpha$, we introduce a novel rate theory formulation. From the standpoint of change in SRO, four possible reactions can be written:
	\begin{equation}\label{ord1}
		\text{Order} \underset{k_1^{'}}{\stackrel{k_1}{\rightleftharpoons}} \text{Order}
	\end{equation}
	
	\begin{equation}\label{ord2}
		\text{Disorder} \underset{k_2^{'}}{\stackrel{k_2}{\rightleftharpoons}} \text{Order}
	\end{equation}
	
	\begin{equation}\label{ord3}
		\text{Disorder} \underset{k_3^{'}}{\stackrel{k_3}{\rightleftharpoons}} \text{Disorder}
	\end{equation}
	
	\begin{equation}\label{ord4}
		\text{Order} \underset{k_4^{'}}{\stackrel{k_4}{\rightleftharpoons}} \text{Disorder}.
	\end{equation}
	
	From equations \ref{ord1}, \ref{ord2}, \ref{ord3} \& \ref{ord4}, the total rate can be calculated as follows:
	\begin{equation}\label{rate1}
		\begin{split}
			R_{total} = \eta k_1 + \eta k_1^{'} + (1-\eta)k_2 + \eta k_2^{'} +\\
			(1-\eta)k_3 + (1-\eta)k_3^{'} + \eta k_4 + (1-\eta)k_4^{'}
		\end{split}.
	\end{equation}
	
	Equation \ref{rate1} can be simplified by the fact that: $k_1^{'}$ = $k_1$, $k_3^{'}$ = $k_3$, $k_4^{'}$ = $k_2$ \& $k_2^{'}$ = $k_4$.
	\begin{equation}\label{rate2}
		\begin{split}
			R_{total} = 2\eta k_1 + 2(1-\eta)k_2 + 2(1-\eta)k_3 + 2\eta k_4
		\end{split}
	\end{equation}
	
	From the four terms on the right side of equation \ref{rate2}, one of them reorders the system (2(1-$\eta$)k$_2$), one disorders (2$\eta$ k$_4$), and two do not change ordering (2$\eta$ k$_1$, 2(1-$\eta$)k$_3$). Therefore, the efficiency term ($\alpha(\eta,T)$) in equation \ref{model} can be written as
	\begin{equation}\label{alpha}
		\alpha(\eta,T) = \frac{k_2(1-\eta) - k_4\eta}{\eta(k_1 + k_4) + (1-\eta)(k_2 + k_3)}.
	\end{equation}
	
	The next step is to find the rates of reactions, $k_1$, $k_2$, $k_3$ \& $k_4$. These rates can be calculated from a Boltzmann equation with barriers obtained from \texttt{TAMMBER} and MD simulations:
	\begin{equation}\label{ord-ord}
		k_1 = \nu_1\exp\left(-\frac{E_{Ord-Ord}}{\rm k_BT}\right),
	\end{equation}
	
	\begin{equation}\label{dis-ord}
		k_2 = \nu_2\exp\left(-\frac{E_{Dis-Ord}}{\rm k_BT}\right),
	\end{equation}
	
	\begin{equation}\label{dis-dis}
		k_3 = \nu_3\exp\left(-\frac{E_{Dis-Dis}}{\rm k_BT}\right),
	\end{equation}
	
	\begin{equation}\label{ord-dis}
		k_4 = \nu_4\exp\left(-\frac{E_{Ord-Dis}}{\rm k_BT}\right)
	\end{equation}
	
	where $\nu_{1-4}$ are pre-exponential factors, and can in principle be extracted from both the \texttt{TAMMBER} and MD results, though for simplicity we take a constant value. Three of these barriers--E$_{Dis-Ord}$, E$_{Dis-Dis}$ \& E$_{Ord-Dis}$--are obtained from \texttt{TAMMBER} as shown in Fig. \ref{migrate}(b). The fourth jump barrier, E$_{Ord-Ord}$, like E$_{Dis-Dis}$ does not change SRO. Therefore, E$_{Ord-Ord}$ was chosen to be equal to Q$_{eff}$ for 100\% LRO from Fig. \ref{diff_active}(b).
	
	The final term in equation \ref{model} that needs to be specified is the jump rate $\frac{dj}{dt}$, which we model as
	\begin{equation}\label{jump_rate}
		\frac{dj}{dt} = C
		\exp\left(-\frac{Q_{eff}}{\rm k_BT}\right),
	\end{equation}
	
	where C is a fitting constant, and Q$_{eff}$ is the total effective migration barrier from Fig. \ref{diff_active}(b). As can be seen from Fig. \ref{diff_active}(b), the total Q$_{eff}$ changes as a function of order parameter. However, it is nearly constant  up to 64\% LRO. Hence, the mean values of Q$_{eff}$, 0.561eV,  is used in equation \ref{jump_rate}.
	
	It should be noted that each of the parameters in the reordering model has a range as shown in table \ref{parameter}. Therefore, instead of just using the mean value, an upper and lower bound with 68\% confidence interval (1$\sigma$) is considered.
	
	\begin{table*}[!ht]
		\caption{Input parameters of the analytical mean-field reordering model}
		\begin{tabular}{ | c | c| c | c | c | }
			\hline
			Parameter & Mean value (eV) & Standard error (eV) & Upper bound (1$\sigma$) (eV) & Lower bound (1$\sigma$) (eV) \\ 
			\hline
			E$_{Ord-Ord}$ & 0.110 & 0.002 & 0.112 & 0.108 \\
			\hline
			E$_{Dis-Ord}$ & 0.535 & 0.016 & 0.551 & 0.519 \\
			\hline
			E$_{Dis-Dis}$ & 0.504 & 0.012 & 0.516 & 0.492 \\
			\hline
			E$_{Ord-Dis}$ & 0.806 & 0.014 & 0.82 & 0.792 \\
			\hline
			Q$_{eff}$ & 0.561 & 0.022 & 0.583 & 0.539 \\
			\hline
		\end{tabular}
		\label{parameter}
	\end{table*}
	
\begin{figure}[!ht]
		\centering
		\includegraphics[scale=0.5]{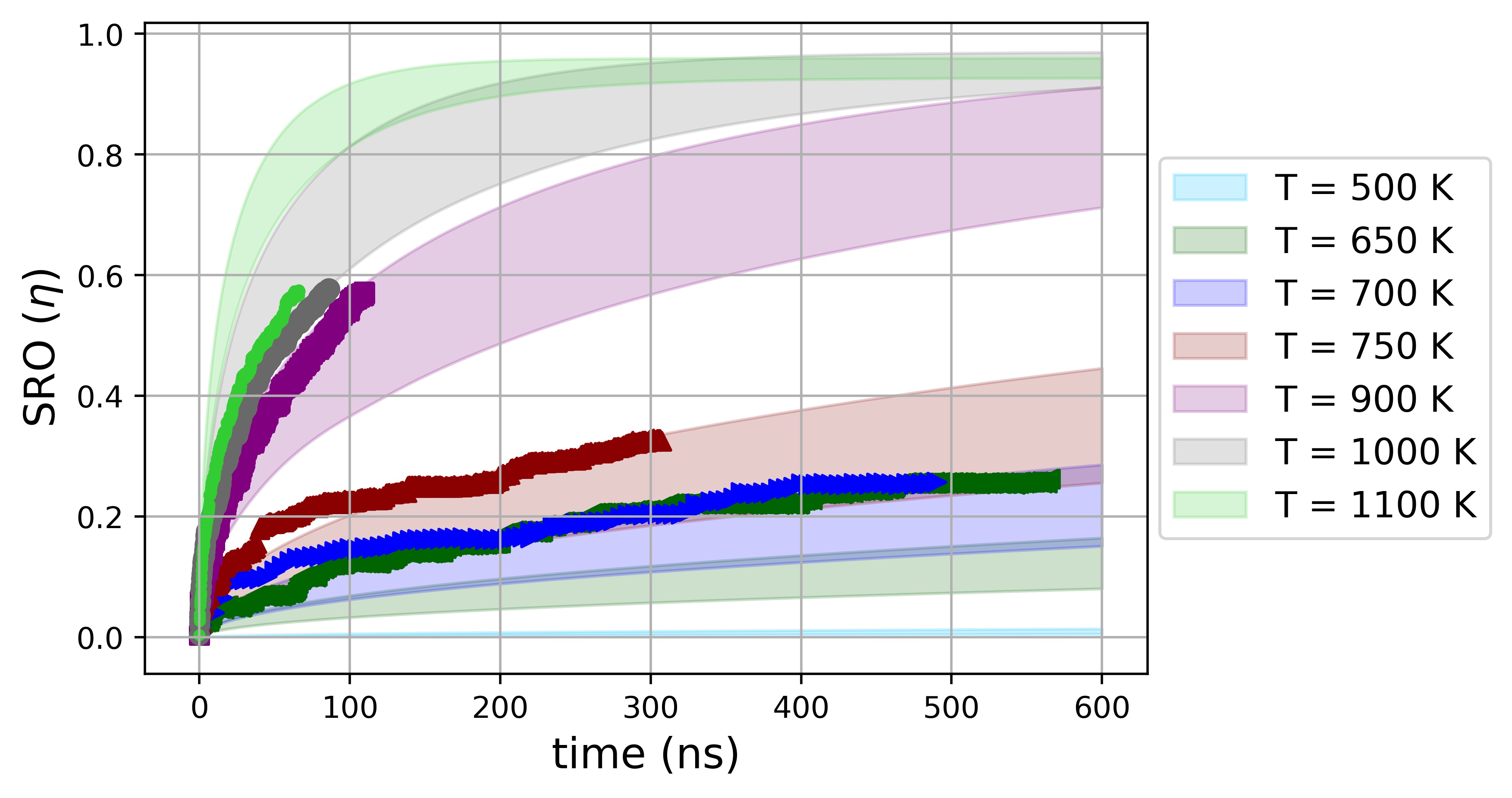}
		\caption{Analytical mean-field model predictions of SRO parameter as a function of time compared to MD simulation results. Each band represent the variance due to the uncertainty associated to the underlying model's parameters (the band bounds a 1-$\sigma$ confidence interval). The narrow band at the bottom of graph shows the prediction at 500 K.}
		\label{compare}
	\end{figure}

	Now with all the parameters specified, the model can be applied and compared with MD results as shown in Fig. \ref{compare}. The agreement with the MD is reasonable, and within statistical accuracy. Note that the MD runs themselves have statistical variance, which is apparent in Figs. \ref{pot_ord} \& \ref{diff_active}(a). There is an intrinsic uncertainty in the MD result. One issue is that we cannot obtain better statistics by running longer simulations—as is common practice in MD—because the system is out-of-equilibrium. That is the main reason for discrepancies between our mean field model and the MD results. We also report a prediction at 500K. It shows that the SIA-induced reordering at 500 K is small. Our analytical model is well-suited for use in broader mean-field rate-theories to study reordering in various radiation conditions in a more comprehensive way. It is worth mentioning that during reordering there will be SIA-vacancy recombination and interactions with extended defects that will affect the kinetics. While this is out-of-scope for our study, we note that point defect recombination can be handled with relative ease in rate-theory, even though model parameters such as recombination radius, sink strengths, and clustering need to be determined.
	
	The Q$_{eff}$ for 0-60\% LRO parameters ($\approx$ 0.6 eV) reported in this study is larger than the value mentioned in reference \cite{zhao2021structural} (0.32 eV). We believe that this difference is mainly due to the use of a different EAM potential. Our study is based on the potential described in Ref. \cite{mishin2004atomistic}--which focuses on Ni$_3$Al--while the newer potential \cite{purja2009development} used in Zhao \& Osetsky's work \cite{zhao2021structural} is meant as a more universal model for all Ni-Al compounds. For sake of comparison, we introduced this 0.32 eV Q$_{eff}$ value to the analytical model (these model outputs are available in the supplementary materials). The main trends predicted by our model did not change.
	
	\section{Conclusion}
	
	SIA diffusion in Ni$_3$Al was studied using MD and \texttt{TAMMBER} simulations. The MD and \texttt{TAMMBER} results are used to parameterize an analytical model able to predict reordering kinetics at low temperatures where MD is not effective. A summary of the most important results is as follows:
	
	1. SIAs can be an effective reordering agent at temperatures lower than 750 K. At these low temperatures, vacancies are not effective reordering agents.\newline
	2. The SIA's small migration barrier (10-50\% of vacancy-mechanism) is a necessary but not sufficient condition for it to be an effective reordering agent at low temperatures. Reordering occurs when Al atoms contribute to SIA diffusion.\newline
	3. In highly ordered systems (LRO parameters larger than 75\%), the SIA migration energy decreases as the order parameter increases. However, the corresponding increase in diffusion coefficients does not correlate with a higher reordering rate, because this fast diffusion is associated with Ni-Ni to Ni-Ni jumps, which do not increase order parameter.\newline
	4. An analytical model has been developed to predict the change in SRO as a function of time during SIA diffusion. Input parameters are obtained from molecular statics and molecular dynamics. The model is predictive and in good agreement with the MD results.
	
	\section{Declaration of interest}
	
	None.
	
	\section{Acknowledgments}
	KF, PS, ZY, LKB, and MRD were financially supported by NSERC-UNENE. KF, PS, ZY, LKB, and MRD thank Compute Canada for generous allocation of computer resources.
	TDS gratefully recognizes support from the Agence Nationale de Recherche, via the MEMOPAS project ANR-19-CE46-0006-1. This work was granted access to the HPC resources of IDRIS under the allocation A0090910965 attributed by GENCI.
	
	\section*{References}
	
	\bibliographystyle{unsrt}
	\bibliography{ref}
	
\end{document}


\maketitle
	
	The supplementary material is presented in four sections. In the first section, the relation between LRO and SRO is explained. In the second section, the change in SRO and potential energy plots as a function of initial order parameter are shown. In the third section, the kinetic properties of SIAs at different temperatures are indicated. In the forth section, the SRO model with lower Q$_{eff}$, 0.32 eV, is shown. In the final part, the number of $<$100$>$ dumbbell throughout the simulation are shown at three temperatures (650, 700 \& 750 K).
	
	\section{Long range and short range order parameters}
	
	The Bragg-Williams LRO parameter \cite{hoyt2011phase} can be calculated from the following equation:
	
	\begin{equation}\label{LRO}
		\xi =  \frac{\text{r}_{\text{Al}}-\text{C}_{\text{Al}}}{1-\text{C}_{\text{Al}}} = \frac{\text{r}_{\text{Ni}}-\text{C}_{\text{Ni}}}{1-\text{C}_{\text{Ni}}}
	\end{equation}
	
	where $\xi$ is the LRO, r$_{\text{Al}}$ is the probability of Al atoms occupying the Al sublattice, and C$_{\text{Al}}$ is the atomic fraction of Al in Ni$_3$Al. The same can be applied for Ni atoms as well. As is explained in the methods section, the disordered structures are made from a fully ordered structure by shuffling atoms to satisfy equation \ref{LRO}.
	
\begin{figure*}[!ht]
		\centering
		
		\includegraphics[width=0.49\textwidth]{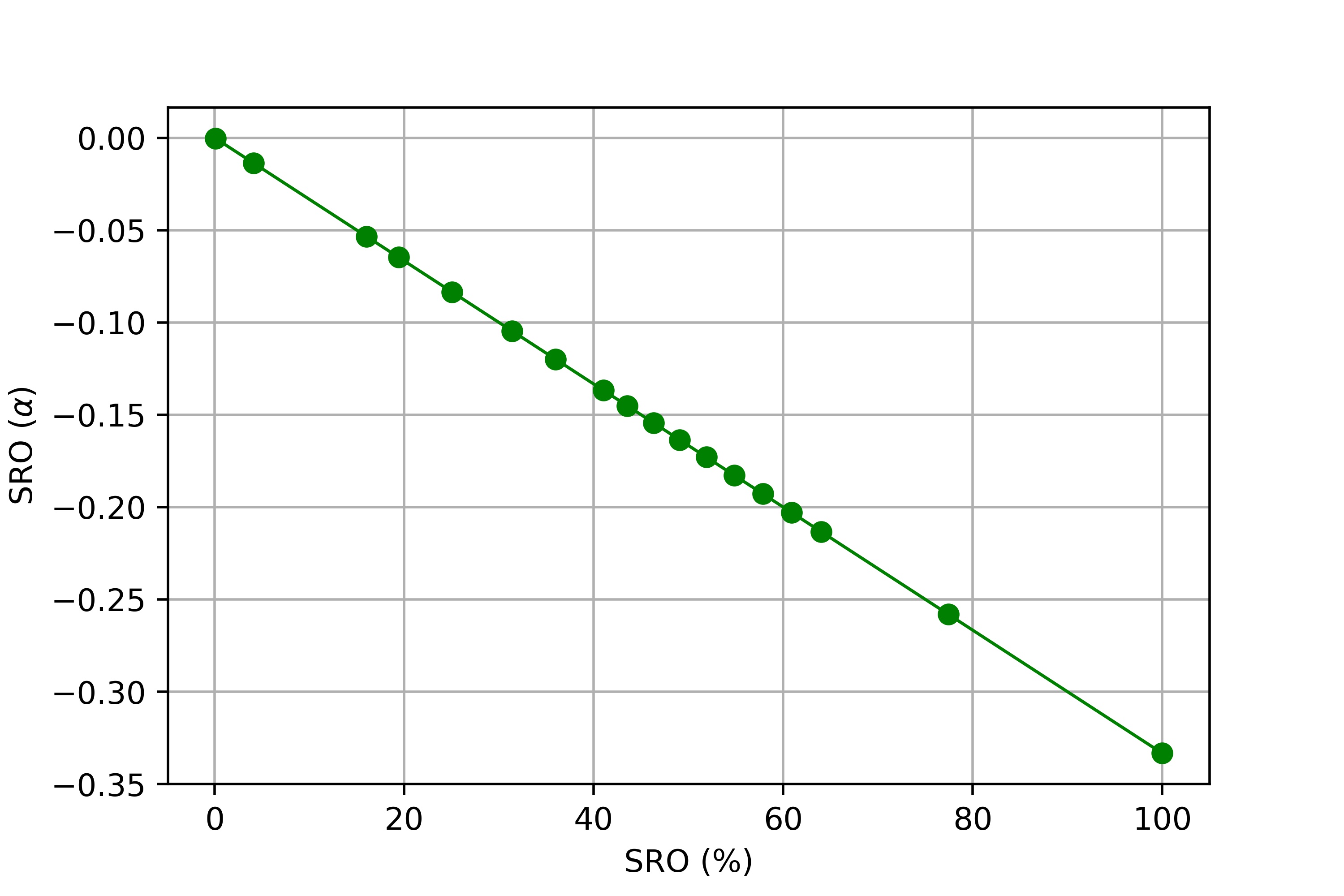}
		\includegraphics[width=0.49\textwidth]{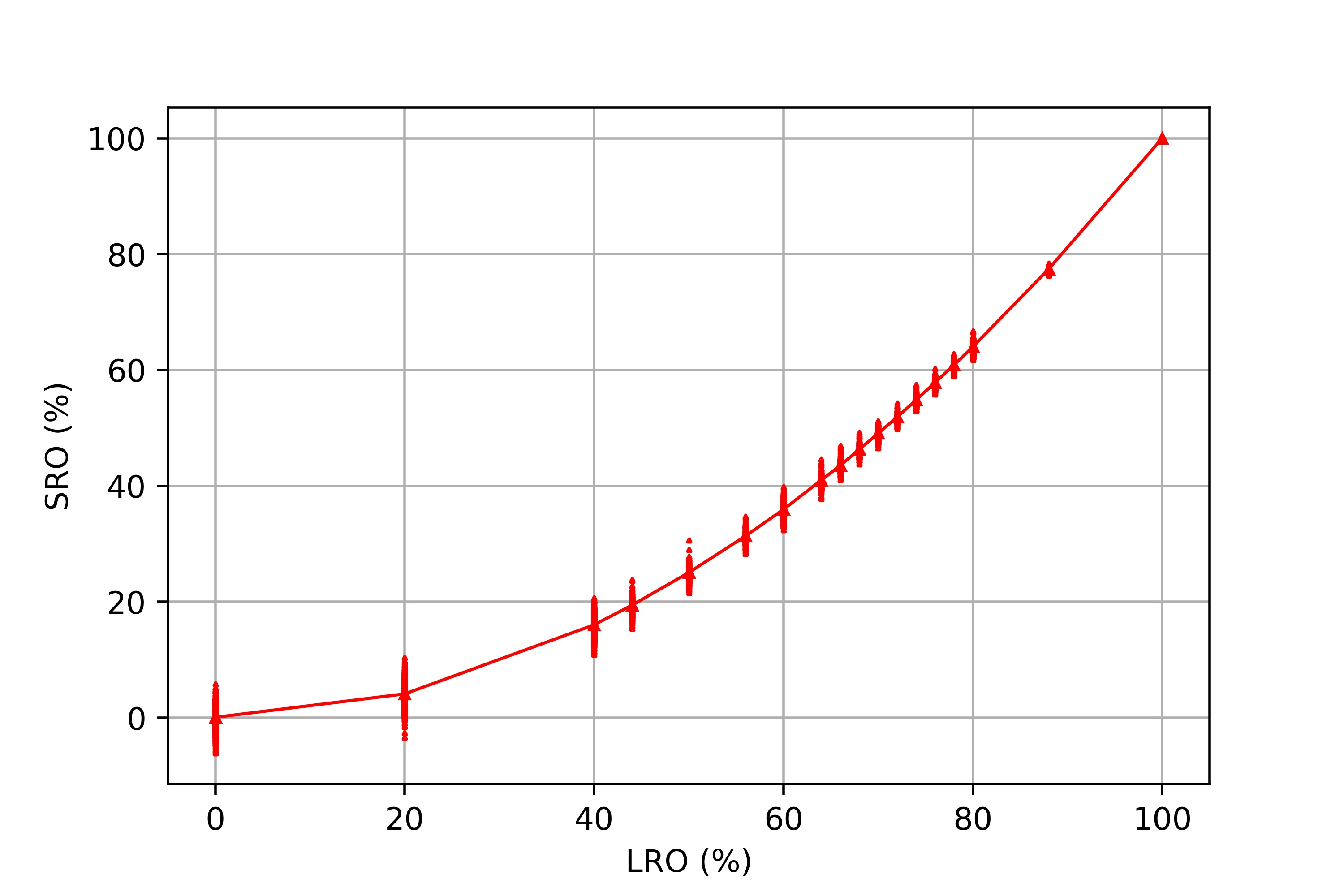}
		\caption{Left: SRO parameter from equations \ref{SRO} and \ref{SRO2}. Right: Relation between LRO and SRO: Each point obtained from an averaged over 1000 configurations (scatter points) with slightly different SROs but the same LRO.}
		\label{LRO-SRO}
	\end{figure*}

	SRO parameter can be calculated using either the equation \ref{SRO} \cite{porter2009phase} as mentioned in the methods section or Warren-Cowley equation (equation \ref{SRO1}) \cite{cowley1950approximate}.
	
	\begin{equation}\label{SRO}
		\text{SRO} =  \frac{\text{P}_{\text{Ni-Al}}-\text{P}^{\text{rand}}_{\text{Ni-Al}}}{\text{P}^{\text{max}}_{\text{Ni-Al}}-\text{P}^{\text{rand}}_{\text{Ni-Al}}}
	\end{equation}
	
	where $\text{P}_{\text{Ni-Al}}$ is the number of Ni-Al bonds. $\text{P}^{\text{max}}_{\text{Ni-Al}}$ is the maximum number of Ni-Al bonds for a fully ordered structure (12000 for a system with 4000 atoms), and $\text{P}^{\text{rand}}_{\text{Ni-Al}}$ is the number of Ni-Al bonds in a fully random disordered structure.
	
	\begin{equation}\label{SRO1}
		\alpha_i =  1 - \frac{N_i}{m_{A}C_{i}}
	\end{equation}
	
	where $\alpha_i$ is the SRO, N$_i$ is the number of A atoms among the C$_i$ atoms in the ith shell, and m$_A$ is the atomic fraction of A. When $\alpha$ is negative, zero, and positive, the system is ordered, fully disordered, and segregated, respectively. This equation can be written again for Ni$_3$Al system, and the number of atoms can be changed by the number of pairs. Considering Al atoms, $\alpha_1$ (first nearest neighbor) can be calculated as following: 
	
	\begin{equation}\label{SRO2}
		\alpha_1 =  1 - \frac{N_{AlNi}}{m_{Ni} N_{tot}}
	\end{equation}
	
	where m$_{Ni}$ is 0.75, N$_{AlNi}$ is the number of Ni around Al or Ni-Al pairs, and N$_{tot}$ is the total neighboring site around Al atoms. For a system with 4000 atoms, N$_{tot}$ is 12000. When the system is fully ordered: N$_{AlNi}$ = 12000, and $\alpha$ = -$\frac{1}{3}$. One can find N$_{AlNi}$ is 9000 for a fully disordered structure, $\alpha=0$. This 9000 is the number Ni-Al bond, $\text{P}^{\text{rand}}_{\text{Ni-Al}}$, used in equation \ref{SRO}. As it is shown in figure \ref{LRO-SRO}(a) either of equations \ref{SRO} and \ref{SRO2} can be used to find SRO.
	
	In order to observe the relation between SRO \& LRO (equations \ref{LRO} and \ref{SRO}), more than 1000 configurations with the same LRO are made. Each of these configurations has a slightly different SRO due to the different number of Ni-Al bonds. The mean and scatter value of SROs are plotted as a function of LROs and shown in figure \ref{LRO-SRO}(b). In this study the initial configurations are made using LRO (equation \ref{LRO}), while all the measurements are done on SRO using equation \ref{SRO}. This is due to a close relation between potential energy and SRO since both of them are strongly depend on the number of Ni-Al bonds.
	
	\section{SRO and potential energies plots}
	
	In the following, the change in SRO and potential energy as a function SD and time at different initial LRO configurations are shown for 650, 700, 750, 900, 1000 \& 1100 K.
	
	\begin{figure*}[!ht]
		\centering
		\includegraphics[scale=0.5]{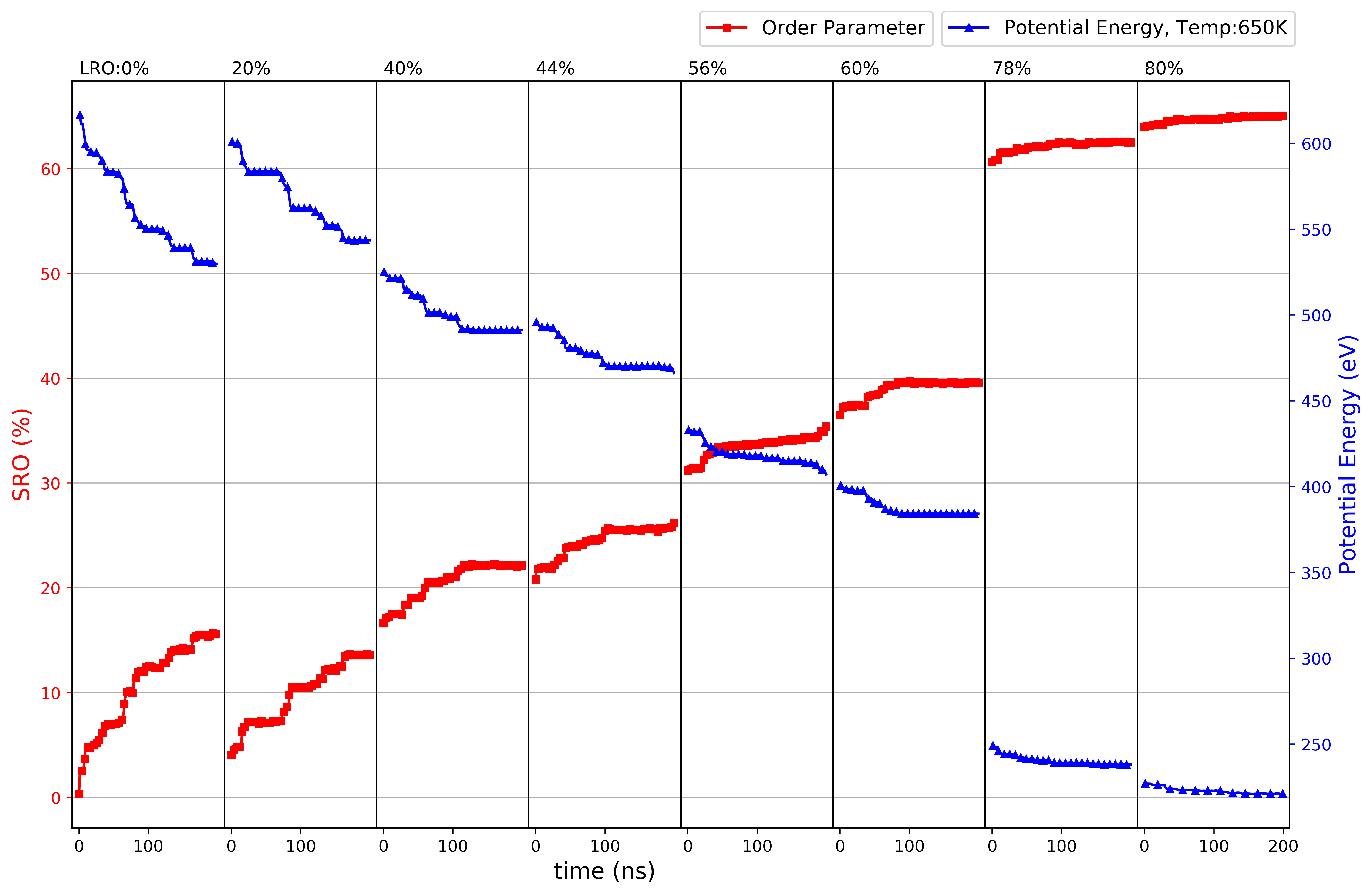}
		\caption{Change in the potential energy and SRO parameter as a function of time at 650 K.}
		\label{pot_ord_1}
	\end{figure*}
	
	\begin{figure*}[!ht]
		\centering
		\includegraphics[scale=0.5]{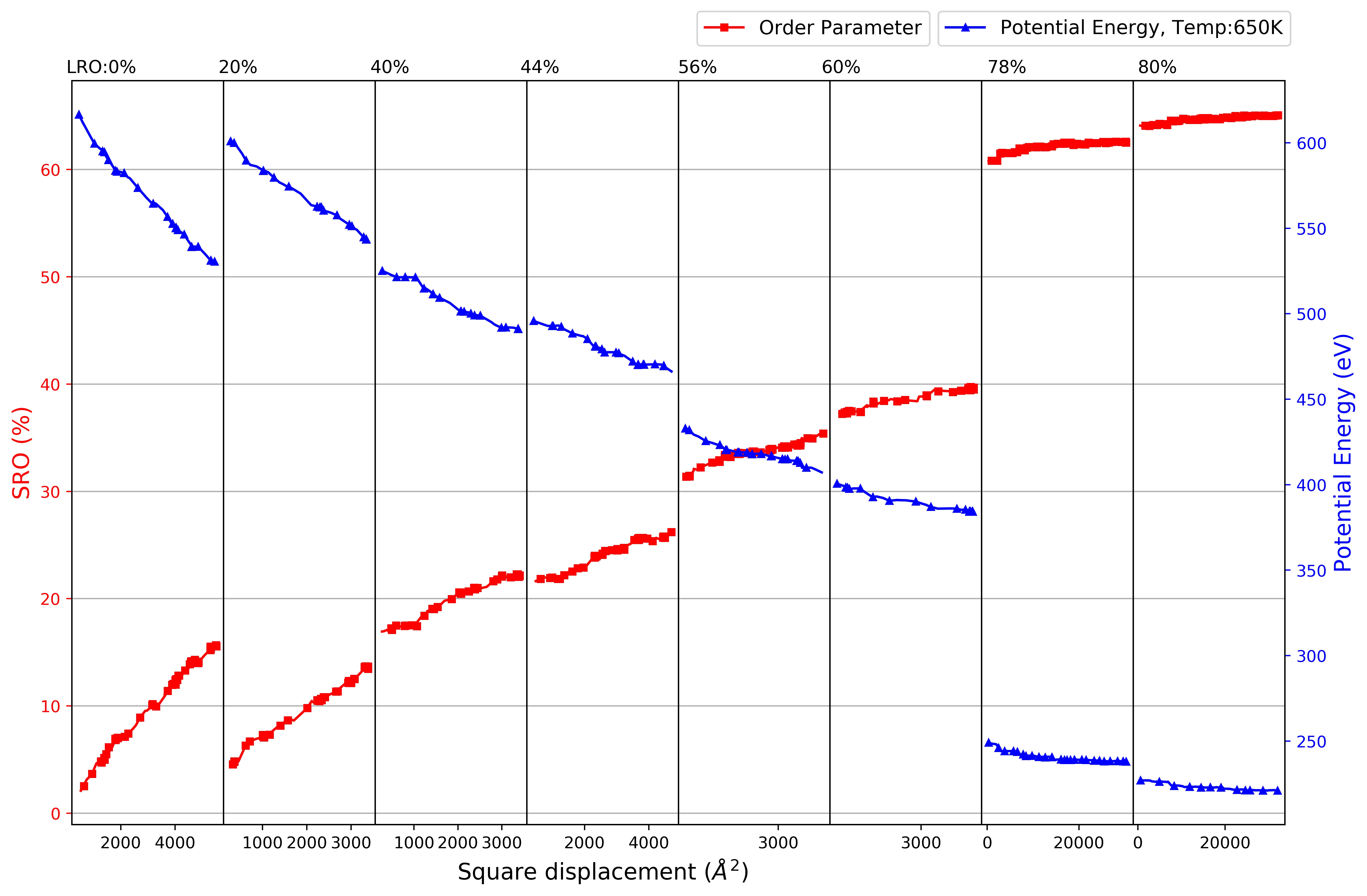}
		\caption{Change in the potential energy and SRO parameter as a function of SD at 650 K.}
		\label{pot_ord_2}
	\end{figure*}
	
	\begin{figure*}[!ht]
		\centering
		\includegraphics[scale=0.7]{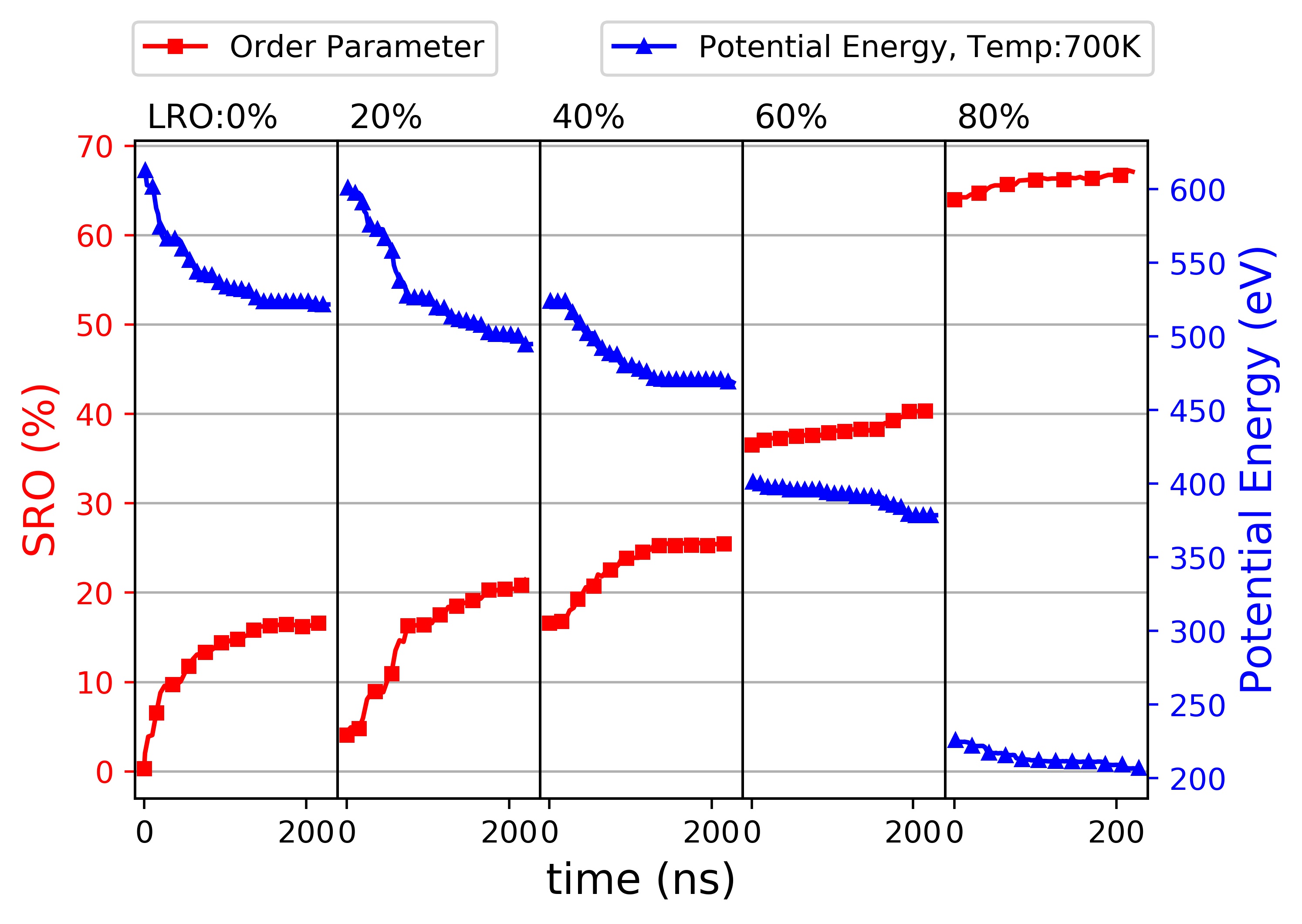}
		\caption{Change in the potential energy and SRO parameter as a function of time at 700 K.}
		\label{pot_ord_3}
	\end{figure*}
	
	\begin{figure*}[!ht]
		\centering
		\includegraphics[scale=0.7]{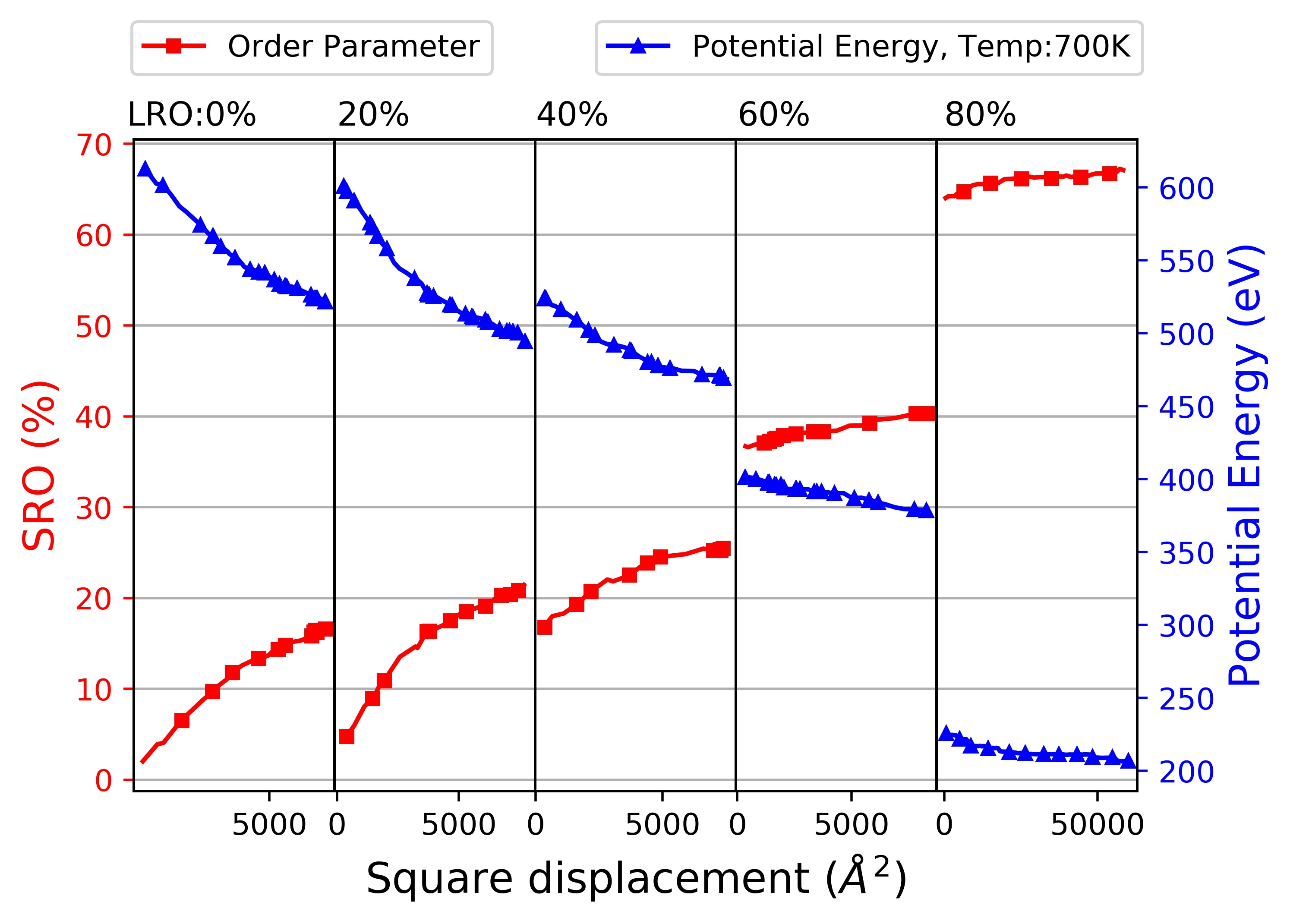}
		\caption{Change in the potential energy and SRO parameter as a function of SD at 700 K.}
		\label{pot_ord_4}
	\end{figure*}
	
	\begin{figure*}[!ht]
		\centering
		\includegraphics[scale=0.5]{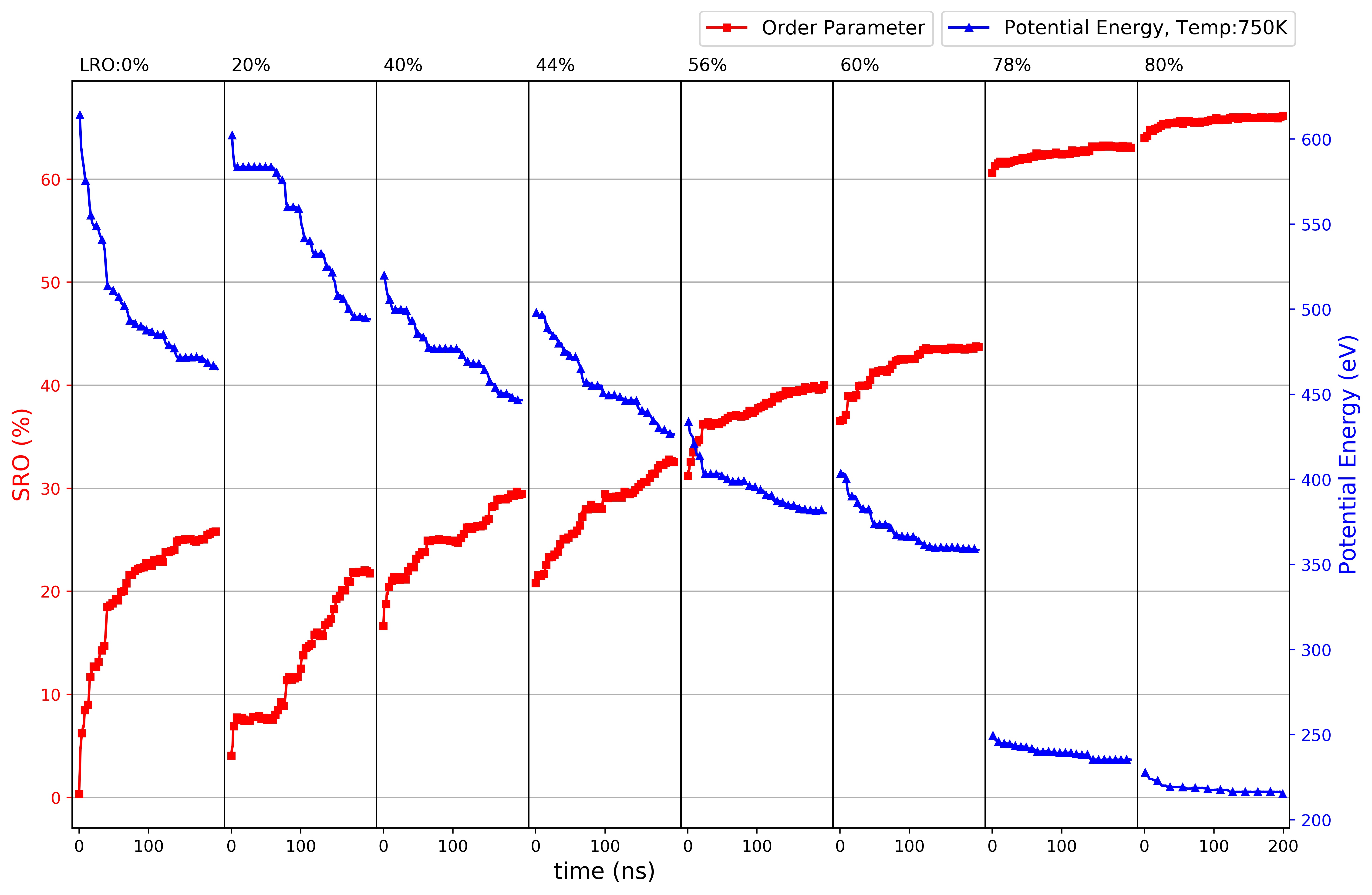}
		\caption{Change in the potential energy and SRO parameter as a function of time at 750 K.}
		\label{pot_ord_5}
	\end{figure*}
	
	\begin{figure*}[!ht]
		\centering
		\includegraphics[scale=0.5]{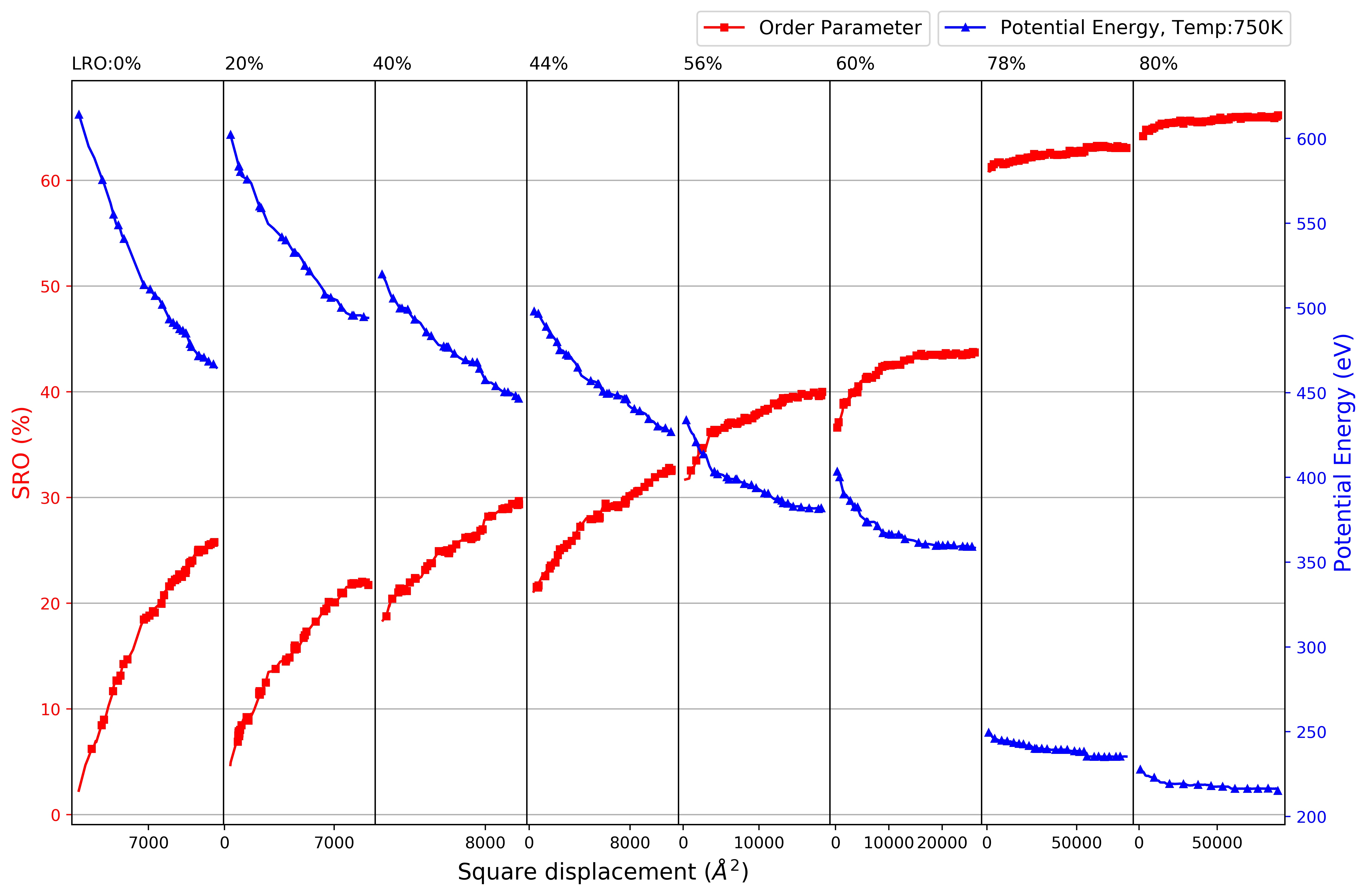}
		\caption{Change in the potential energy and SRO parameter as a function of SD at 750 K.}
		\label{pot_ord_6}
	\end{figure*}
	
	\begin{figure*}[!ht]
		\centering
		\includegraphics[scale=0.4]{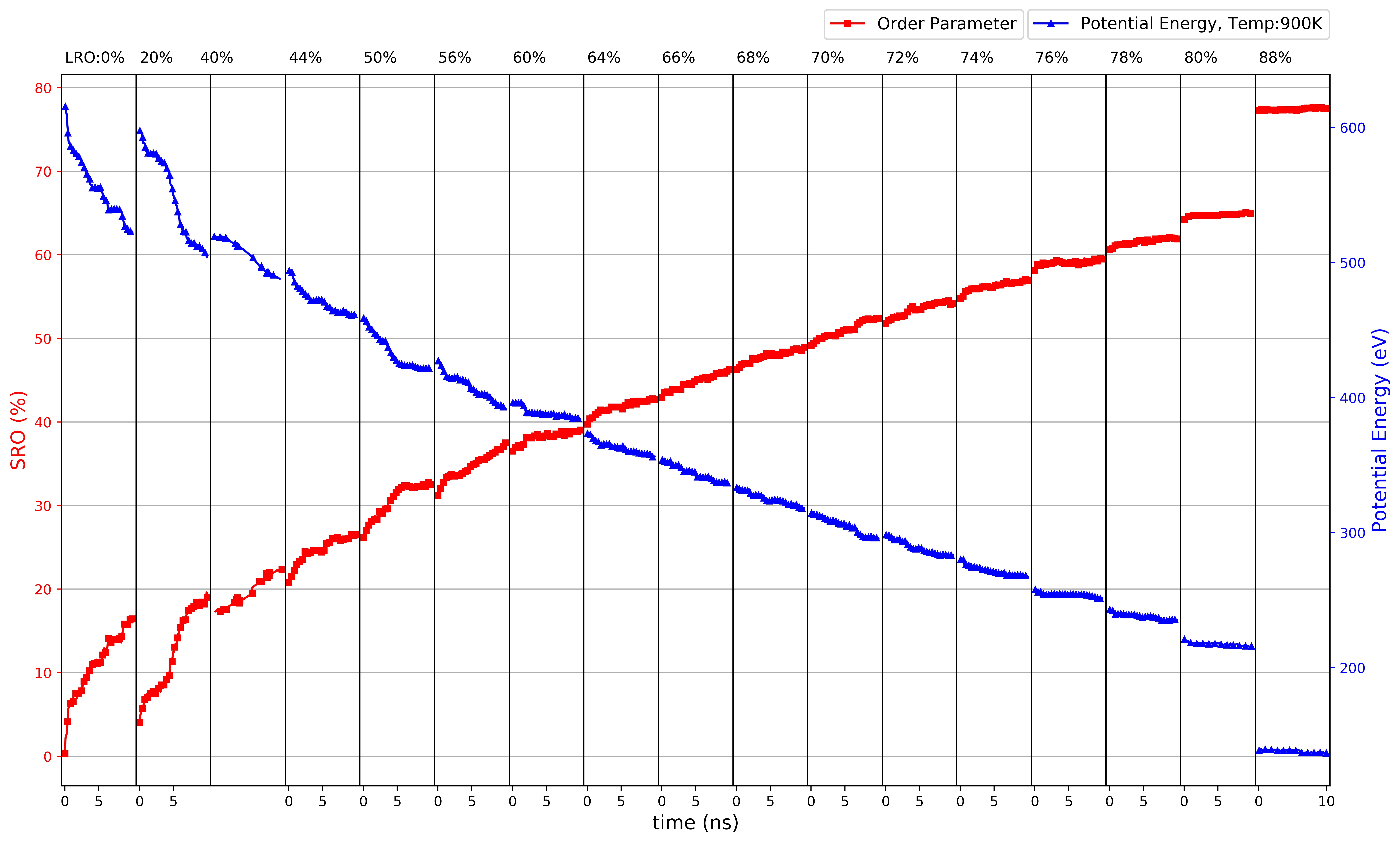}
		\caption{Change in the potential energy and SRO parameter as a function of time at 900 K.}
		\label{pot_ord_7}
	\end{figure*}
	
	\begin{figure*}[!ht]
		\centering
		\includegraphics[scale=0.4]{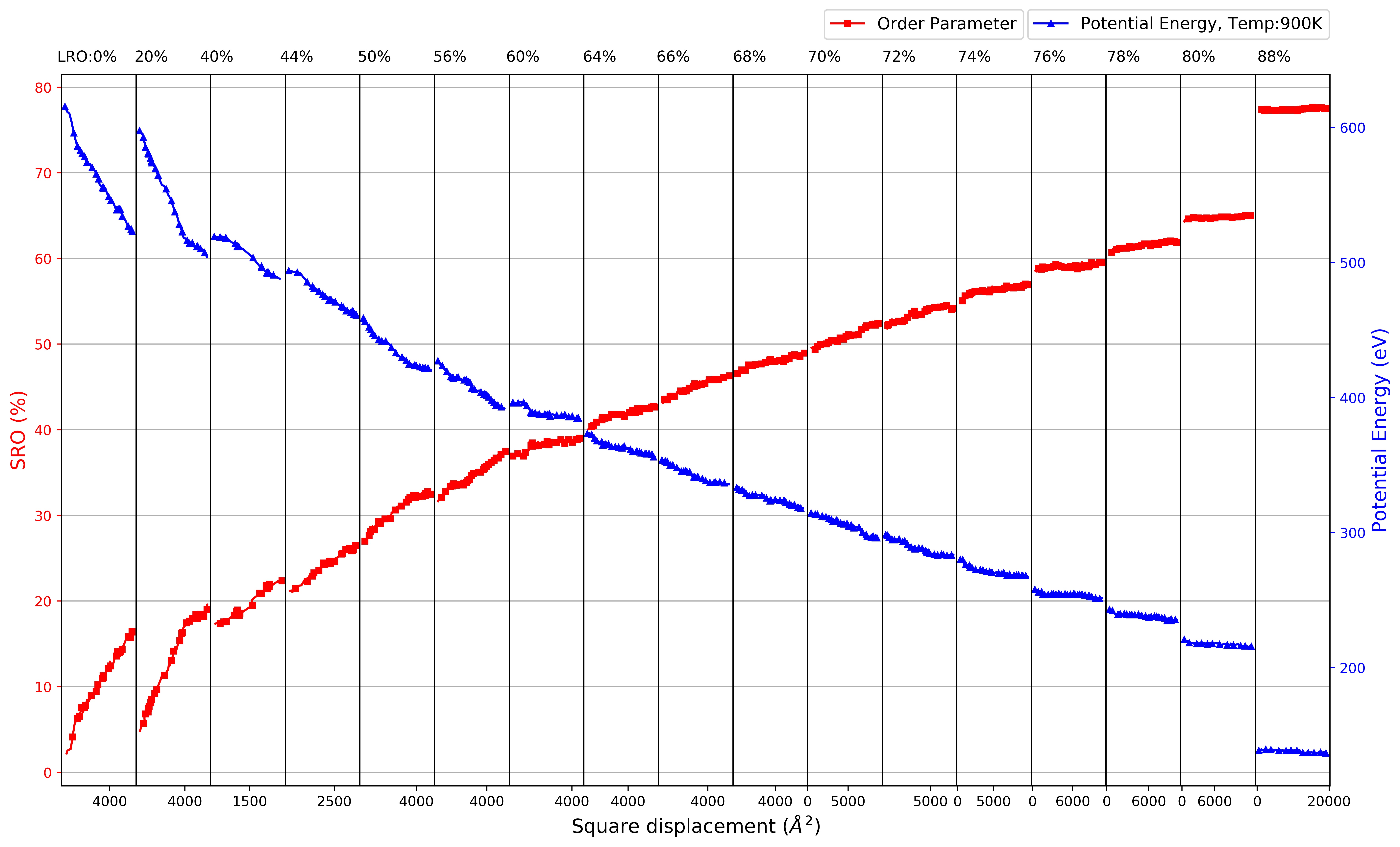}
		\caption{Change in the potential energy and SRO parameter as a function of SD at 900 K.}
		\label{pot_ord_8}
	\end{figure*}
	
	\begin{figure*}[!ht]
		\centering
		\includegraphics[scale=0.4]{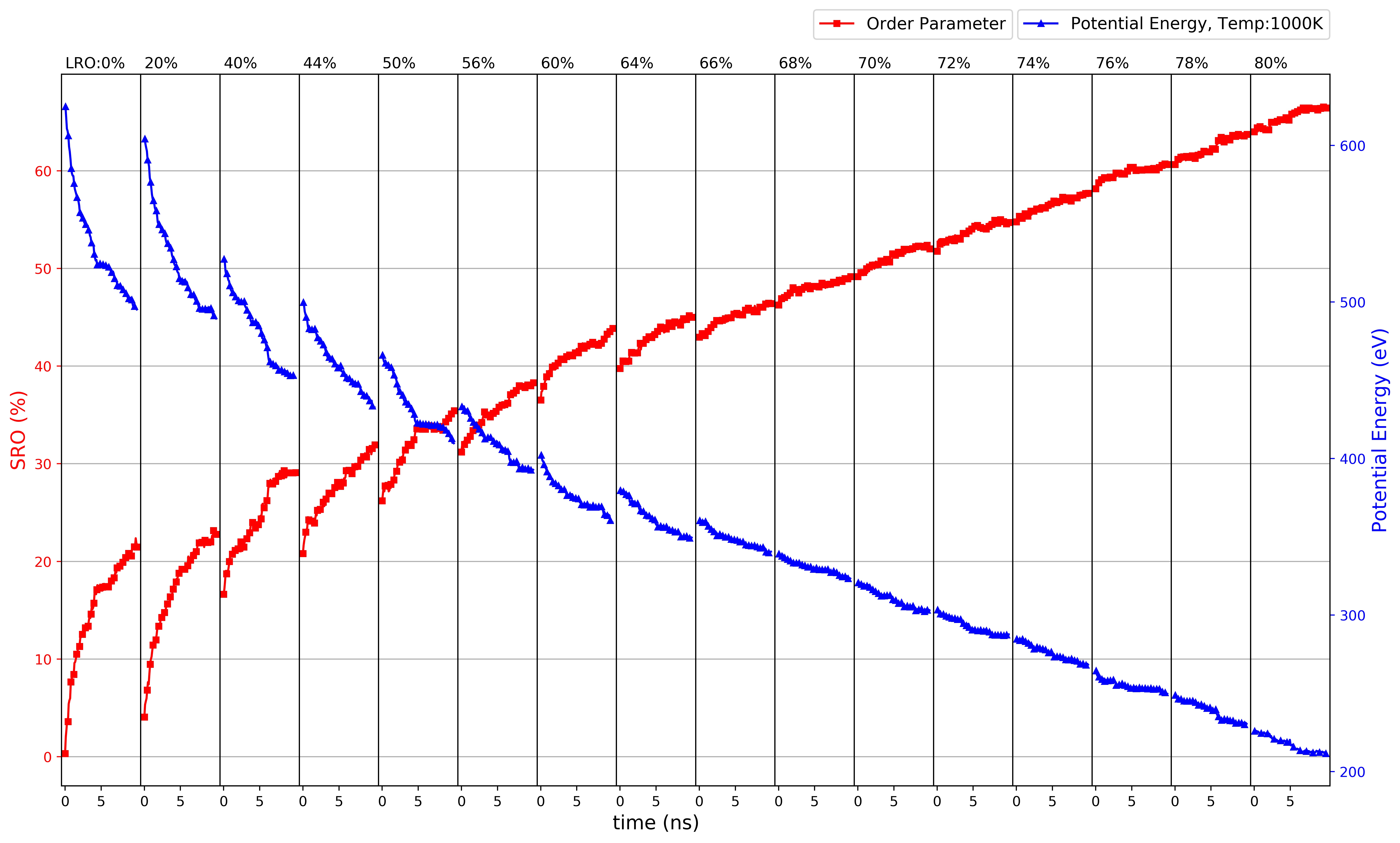}
		\caption{Change in the potential energy and SRO parameter as a function of time at 1000 K.}
		\label{pot_ord_9}
	\end{figure*}
	
	\begin{figure*}[!ht]
		\centering
		\includegraphics[scale=0.4]{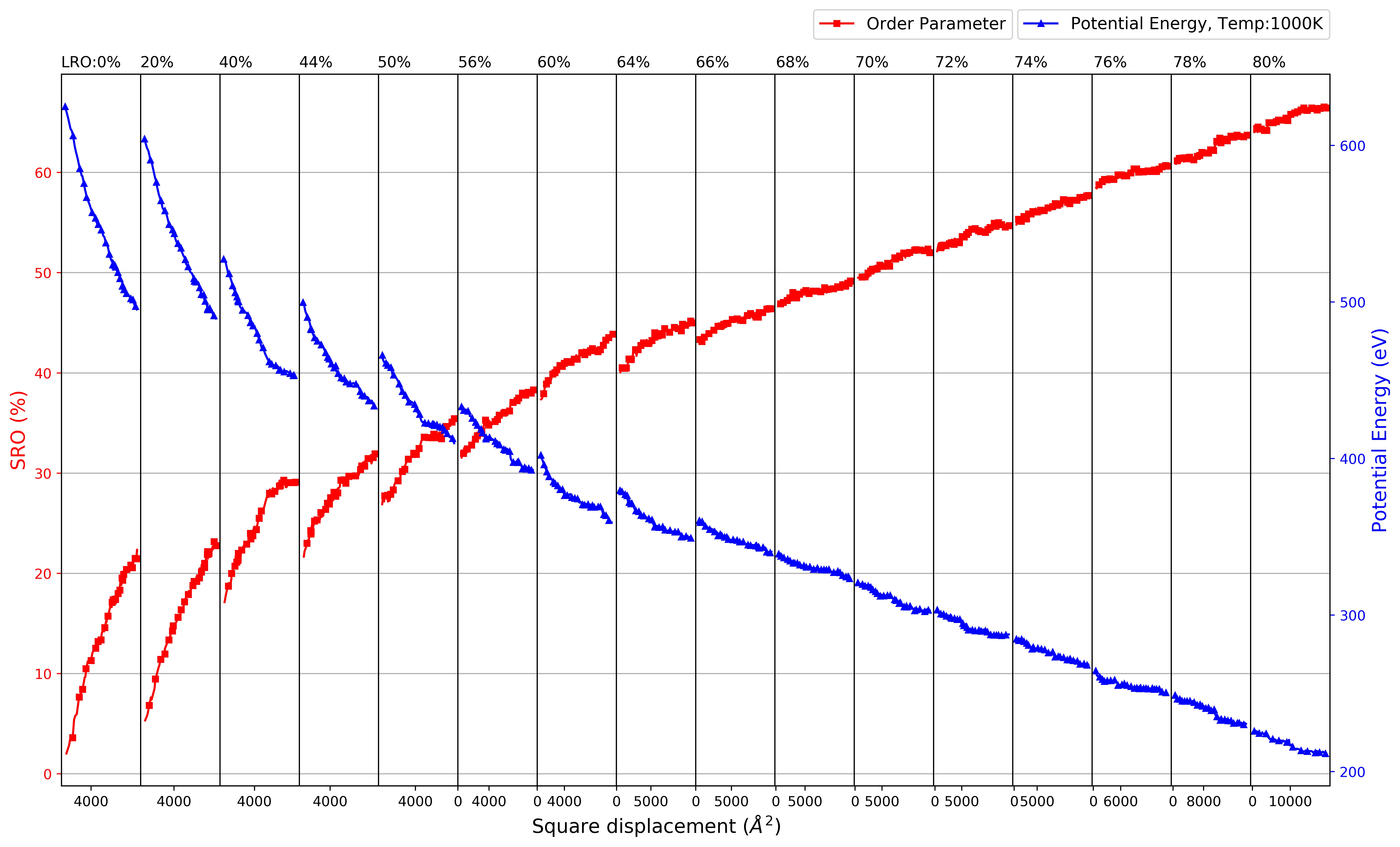}
		\caption{Change in the potential energy and SRO parameter as a function of SD at 1000 K.}
		\label{pot_ord_10}
	\end{figure*}
	
	\begin{figure*}[!ht]
		\centering
		\includegraphics[scale=0.4]{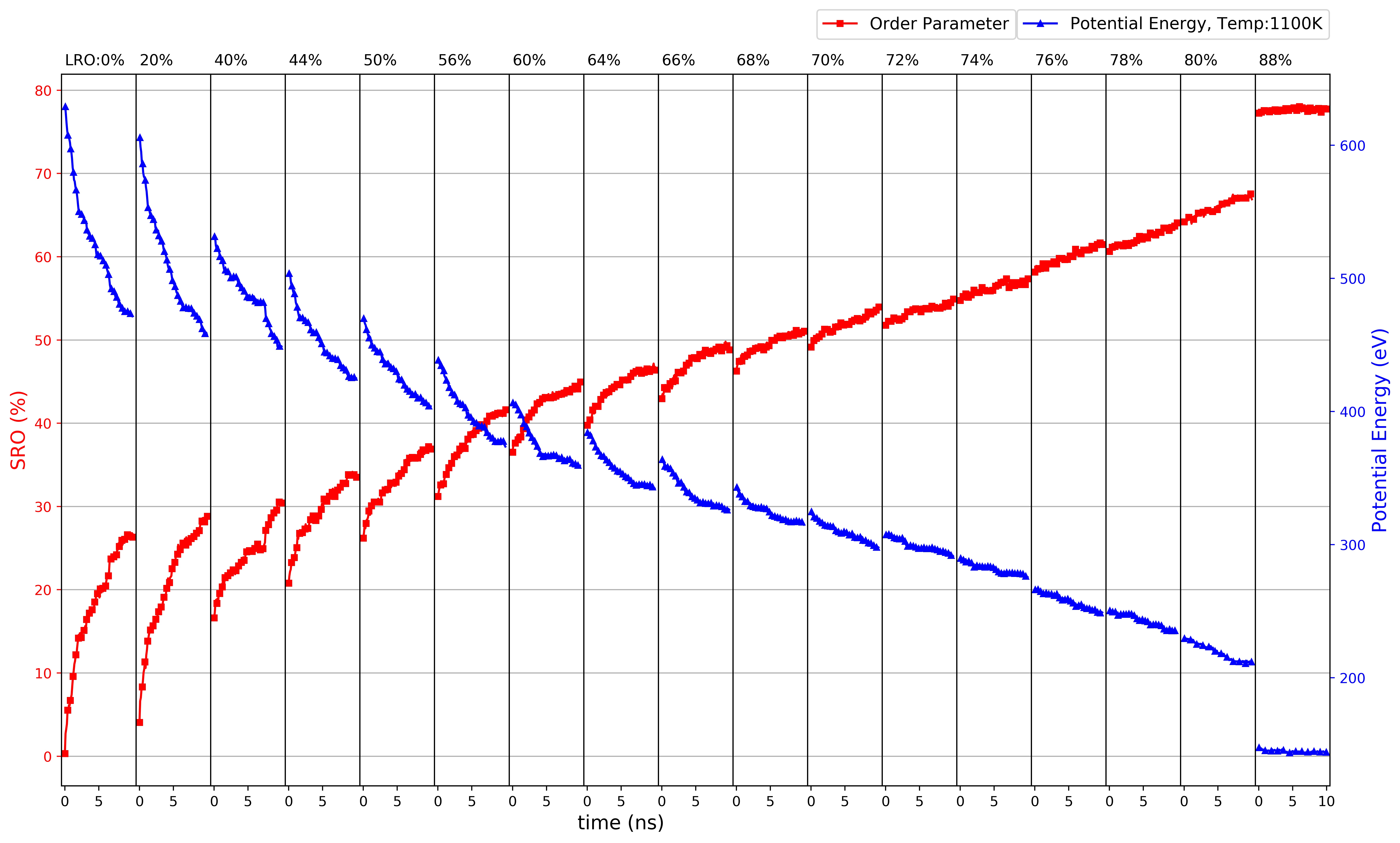}
		\caption{Change in the potential energy and SRO parameter as a function of time at 1100 K.}
		\label{pot_ord_11}
	\end{figure*}
	
	\begin{figure*}[!ht]
		\centering
		\includegraphics[scale=0.4]{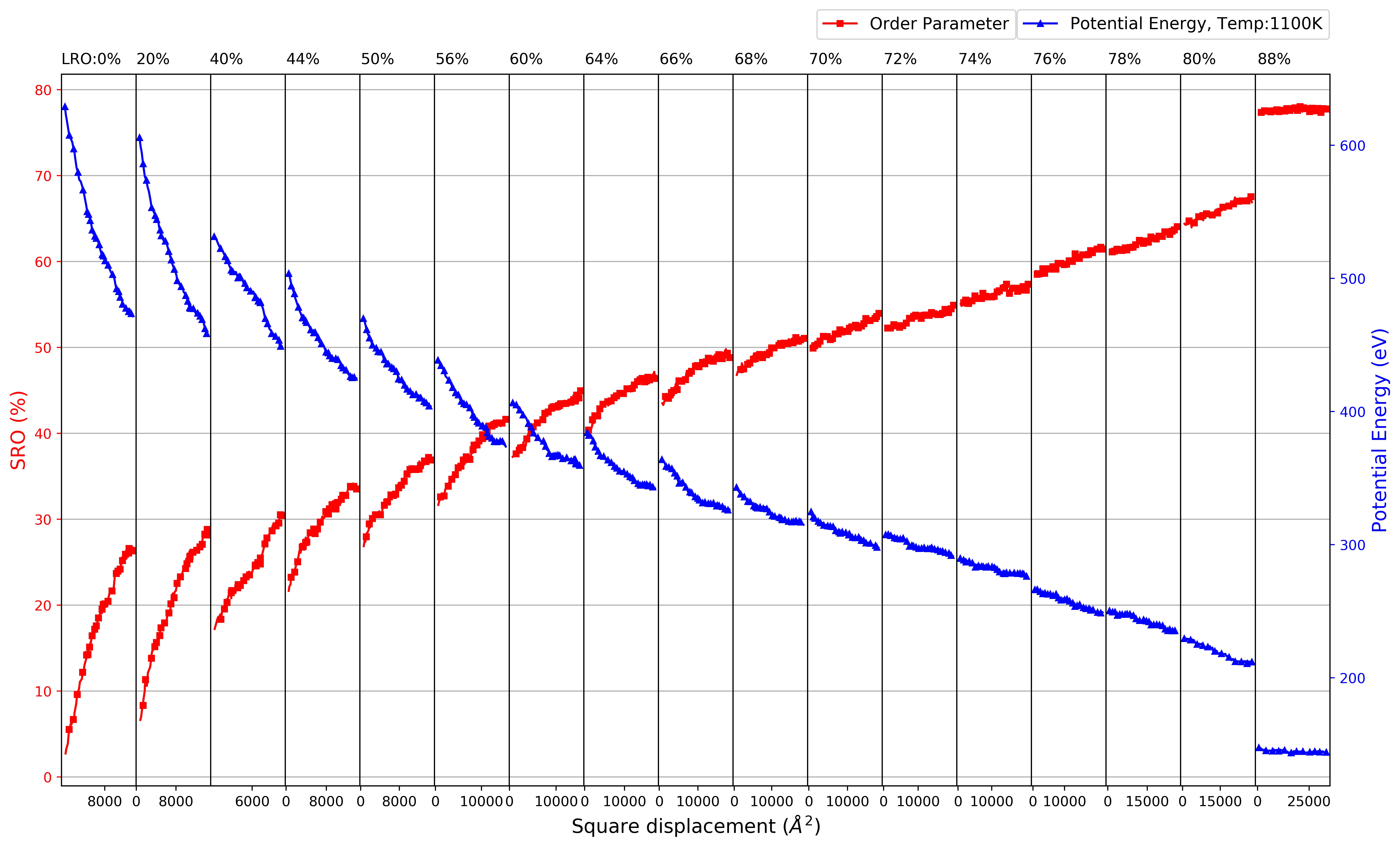}
		\caption{Change in the potential energy and SRO parameter as a function of SD at 1100 K.}
		\label{pot_ord_12}
	\end{figure*}
	
	\section{Kinetic properties of SIA}
	
\begin{figure*}[!ht]
		\centering
		\includegraphics[scale=0.6]{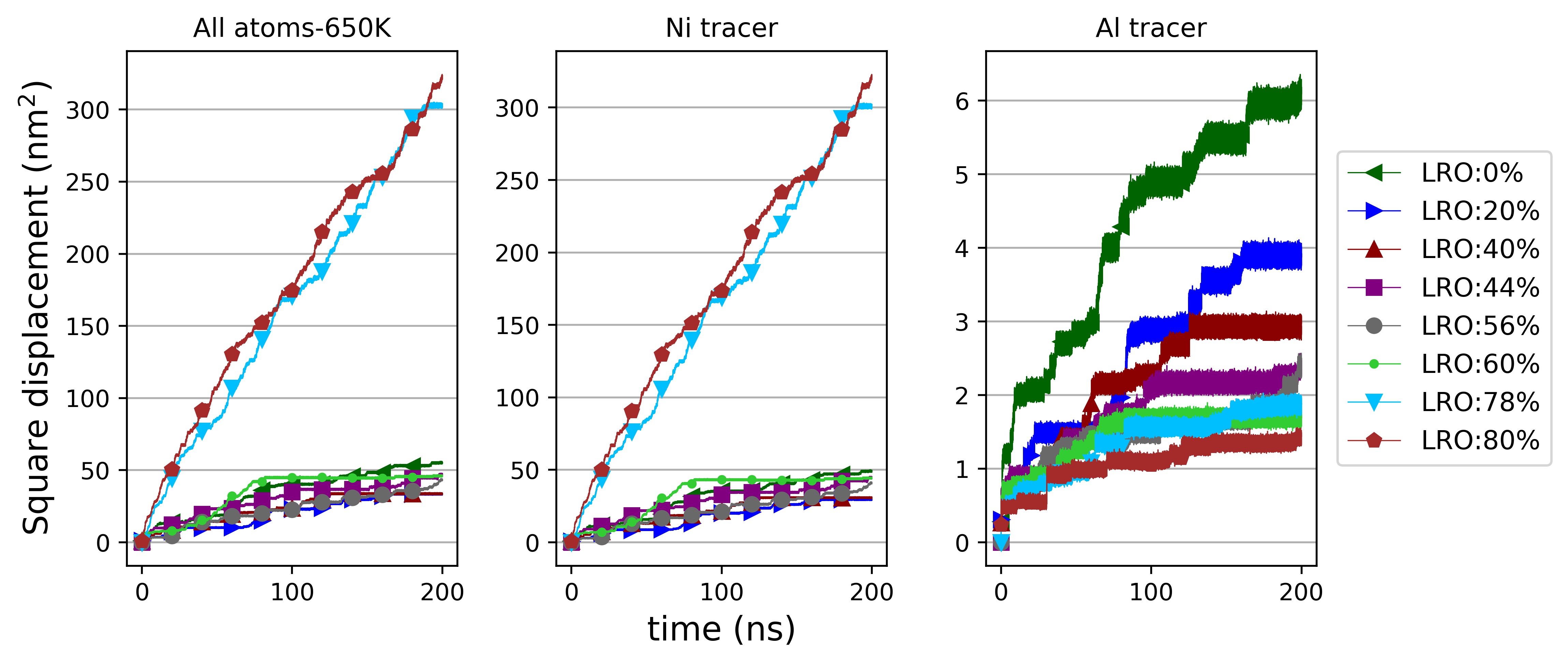}
		\caption{Atomic square displacement as a function of time at 650 K, given different initial configurations (LRO parameters).}
		\label{Square1}
	\end{figure*}

\begin{figure*}[!ht]
		\centering
		\includegraphics[scale=0.6]{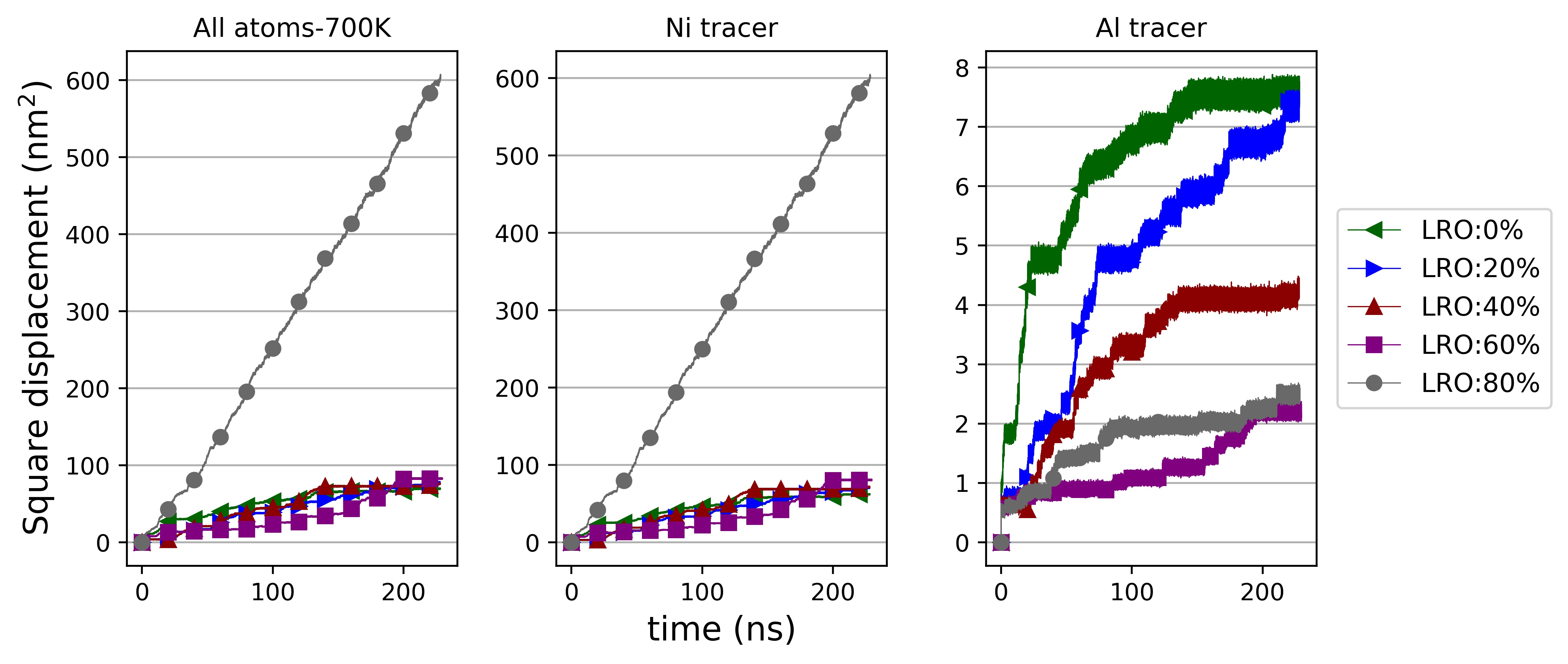}
		\caption{Atomic square displacement as a function of time at 700 K, given different initial configurations (LRO parameters).}
		\label{Square2}
	\end{figure*}

\begin{figure*}[!ht]
		\centering
		\includegraphics[scale=0.6]{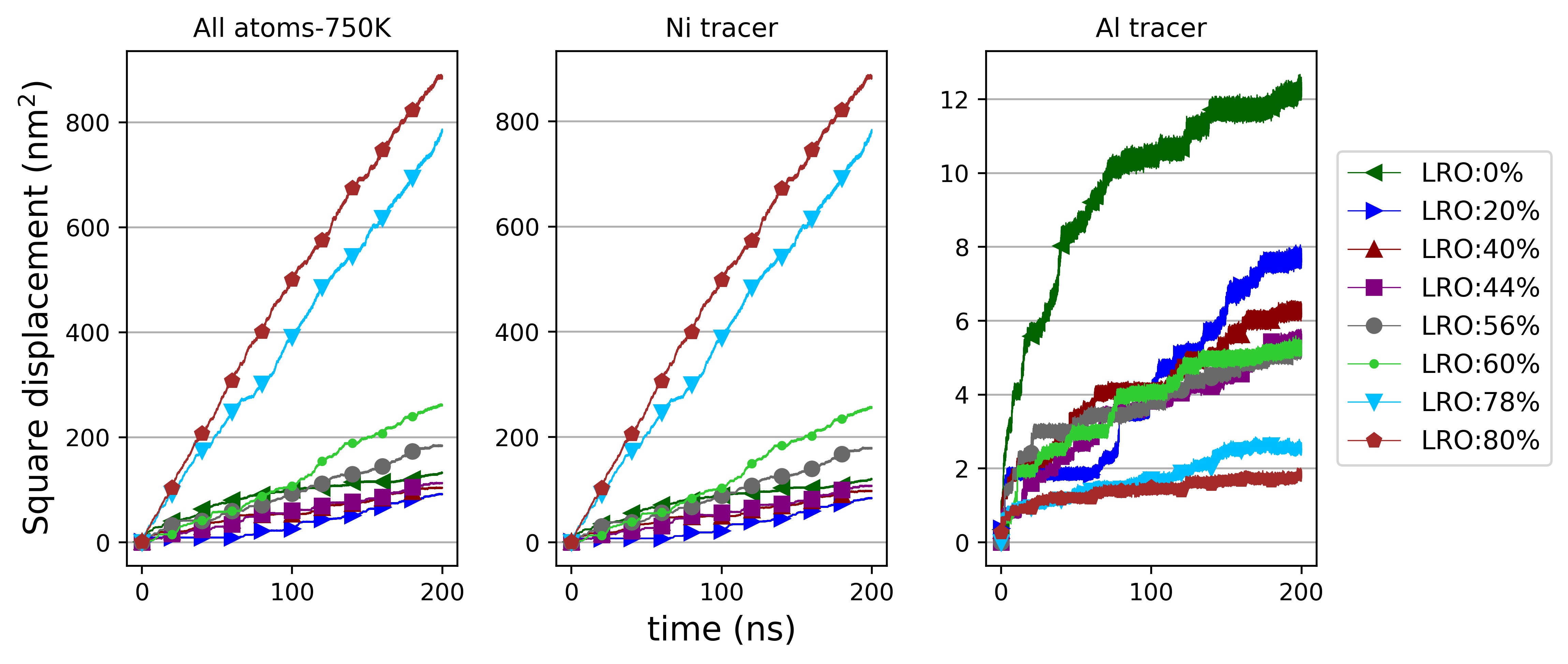}
		\caption{Atomic square displacement as a function of time at 750 K, given different initial configurations (LRO parameters).}
		\label{Square3}
	\end{figure*}

\begin{figure*}[!ht]
		\centering
		\includegraphics[scale=0.6]{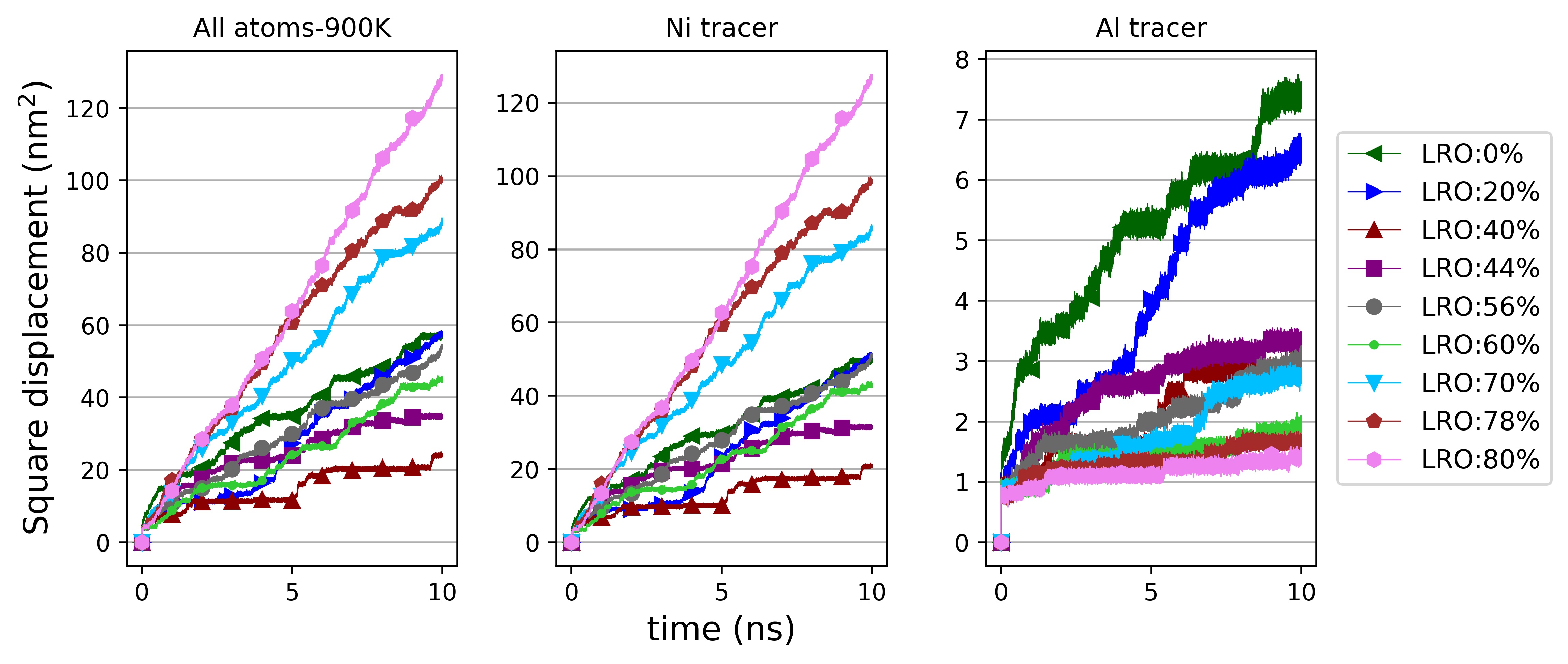}
		\caption{Atomic square displacement as a function of time at 900 K, given different initial configurations (LRO parameters).}
		\label{Square4}
	\end{figure*}

\begin{figure*}[!ht]
		\centering
		\includegraphics[scale=0.6]{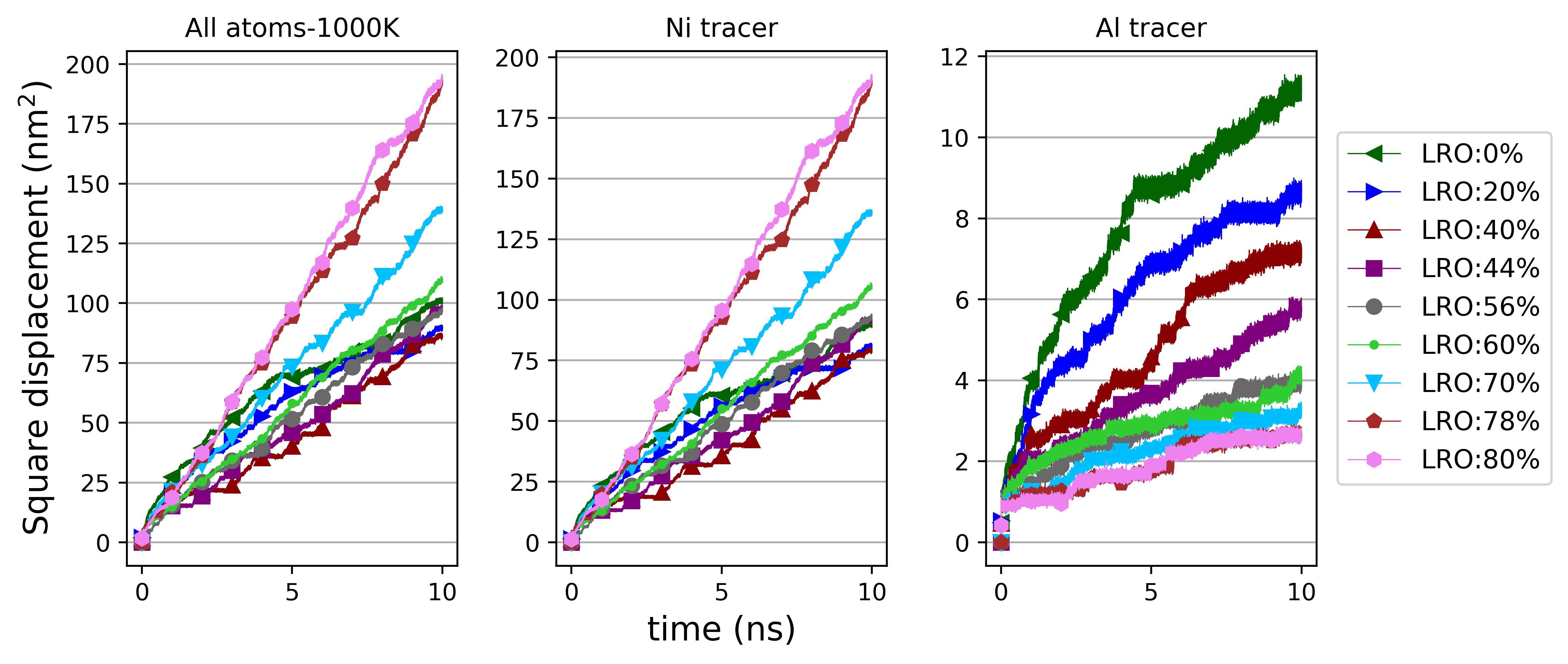}
		\caption{Atomic square displacement as a function of time at 1000 K, given different initial configurations (LRO parameters).}
		\label{Square5}
	\end{figure*}

\begin{figure*}[bht]
		\centering
		\begin{subfigure}[b]{.45\linewidth}
			\includegraphics[scale=0.5]{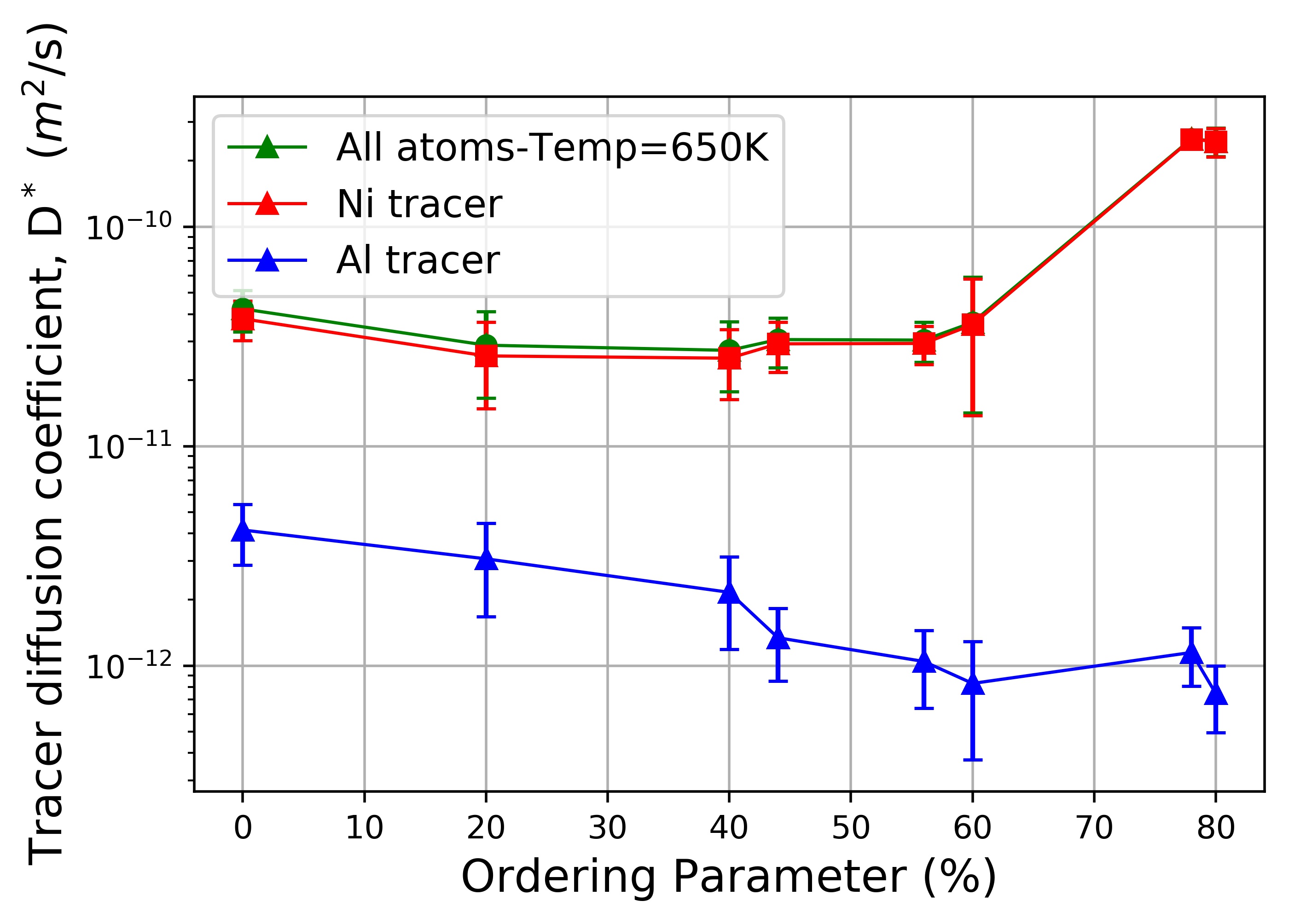}
			\caption{650K}
		\end{subfigure}
		\begin{subfigure}[b]{.45\linewidth}
			\includegraphics[scale=0.5]{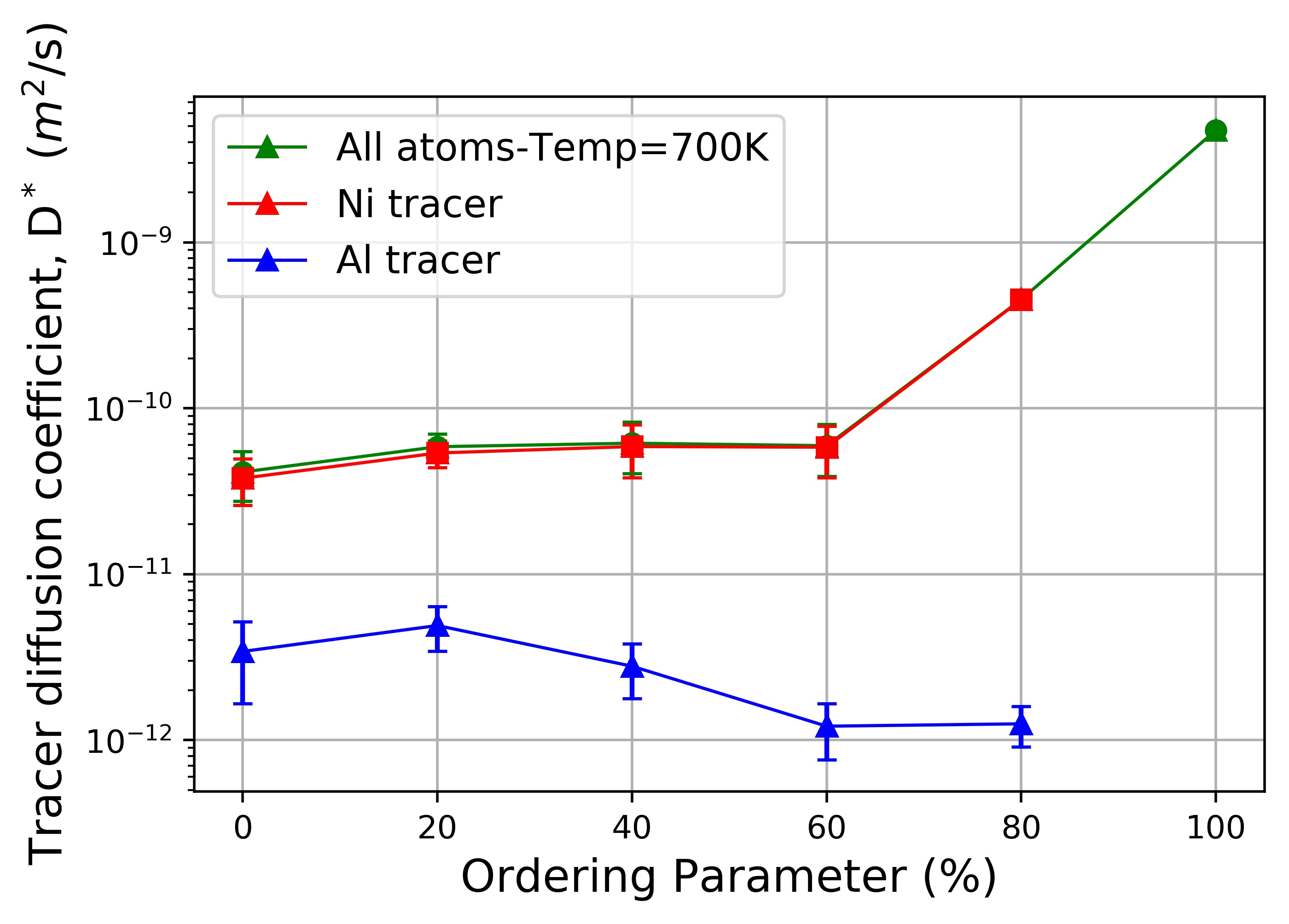}
			\caption{700K}
		\end{subfigure}\\
		\begin{subfigure}[b]{.45\linewidth}
			\includegraphics[scale=0.5]{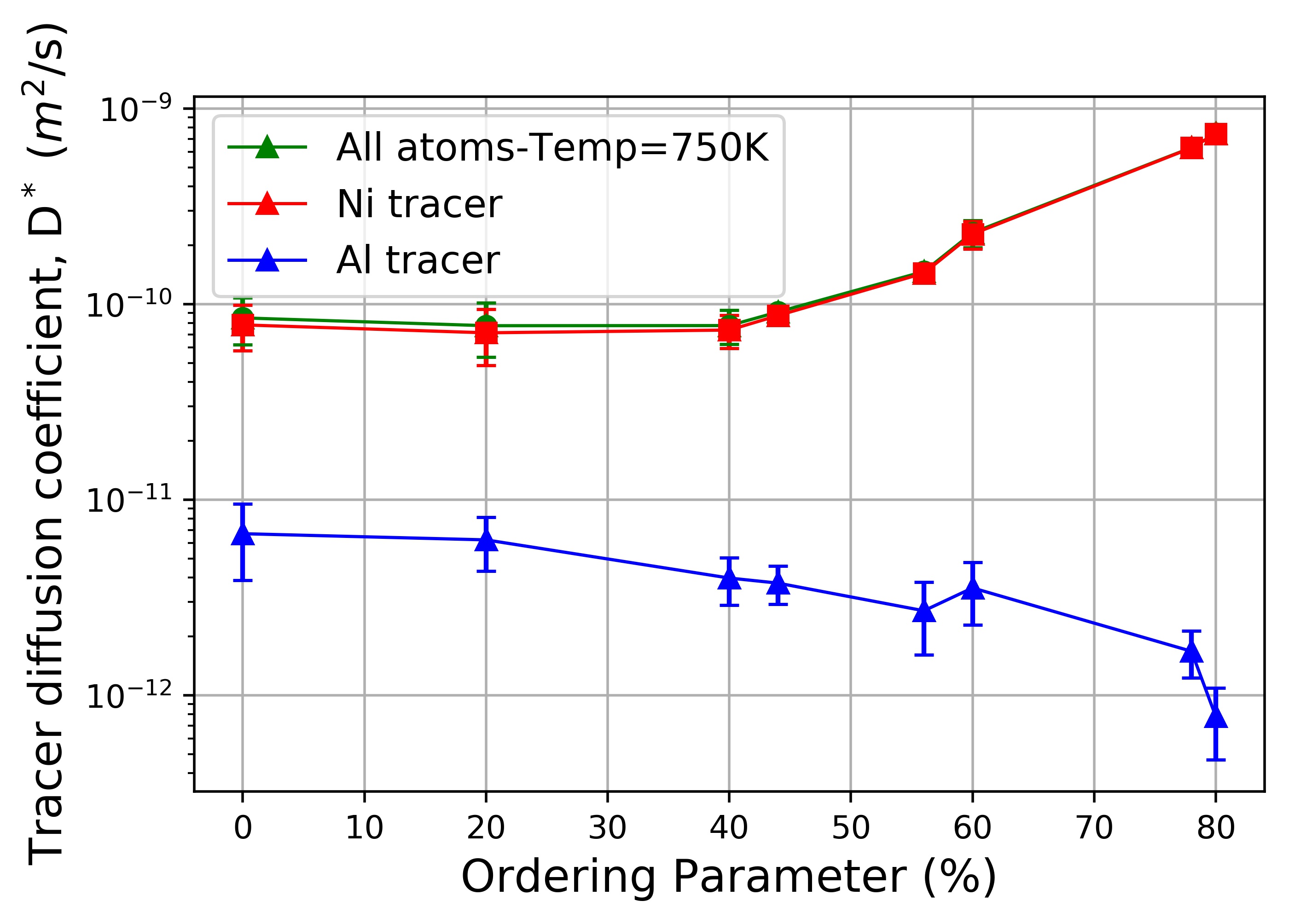}
			\caption{750K}
		\end{subfigure}
		\begin{subfigure}[b]{.45\linewidth}
			\includegraphics[scale=0.5]{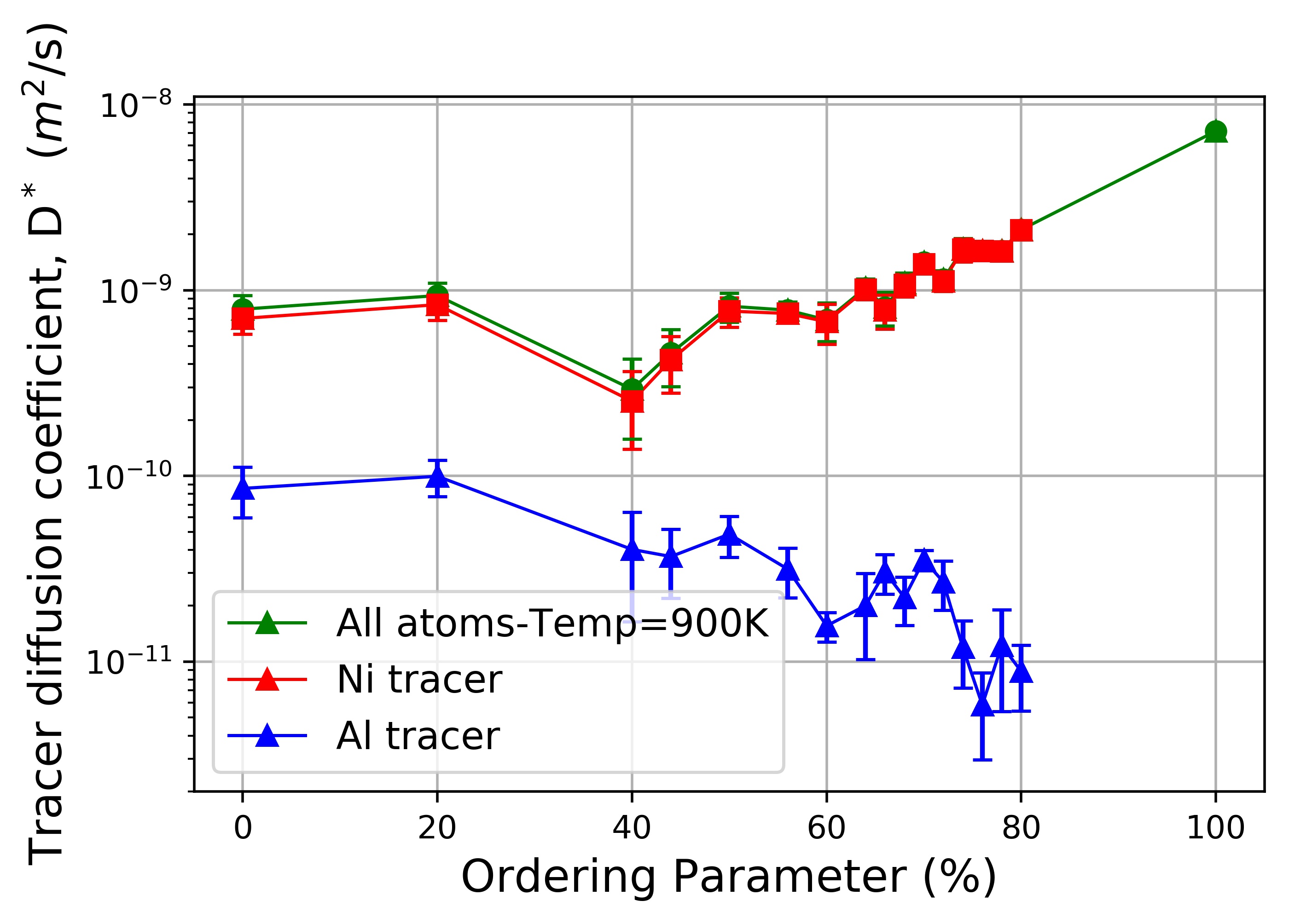}
			\caption{900K}
		\end{subfigure}\\
		\begin{subfigure}[b]{.45\linewidth}
			\includegraphics[scale=0.5]{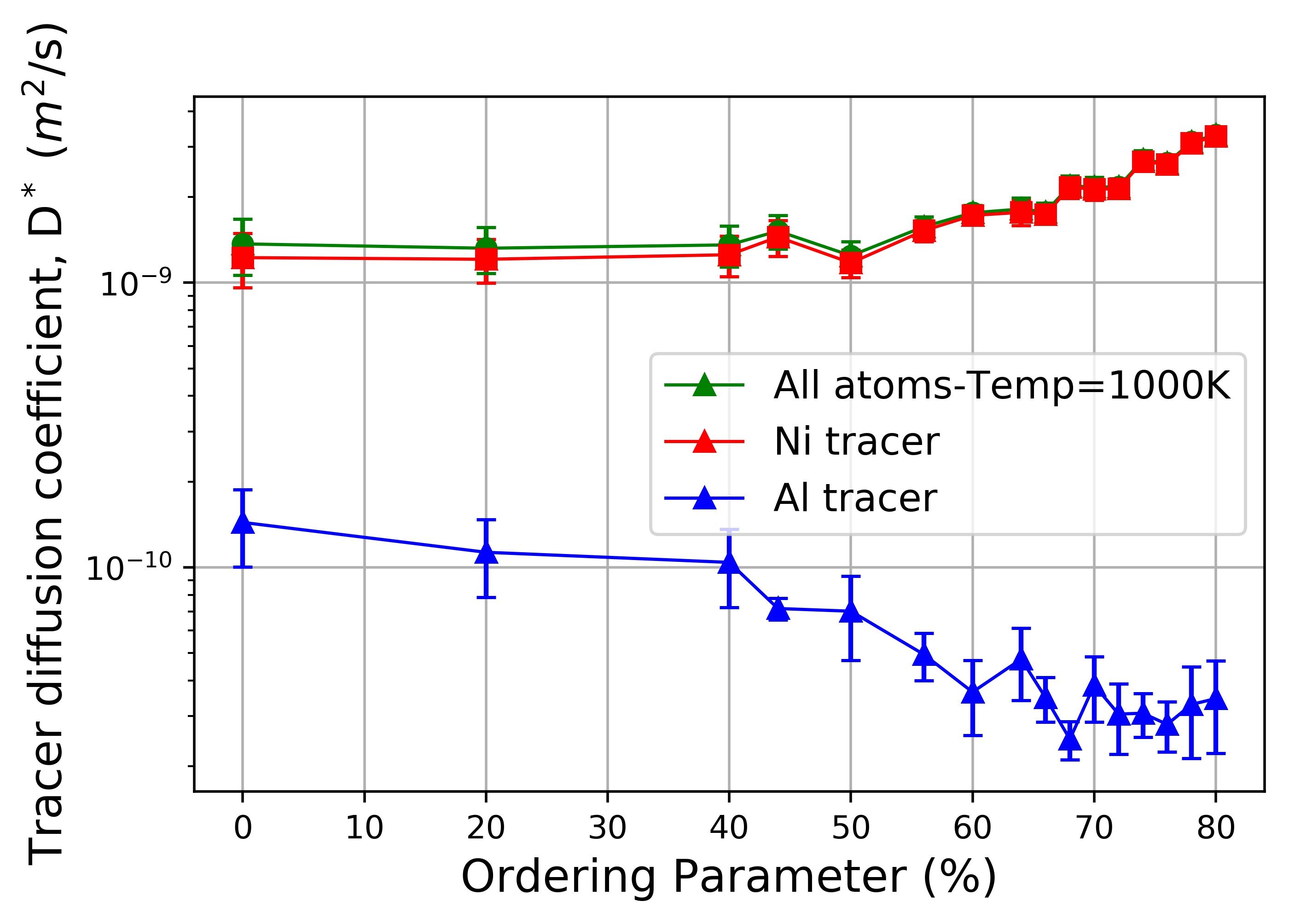}
			\caption{1000K}
		\end{subfigure}
		\begin{subfigure}[b]{.45\linewidth}
			\includegraphics[scale=0.5]{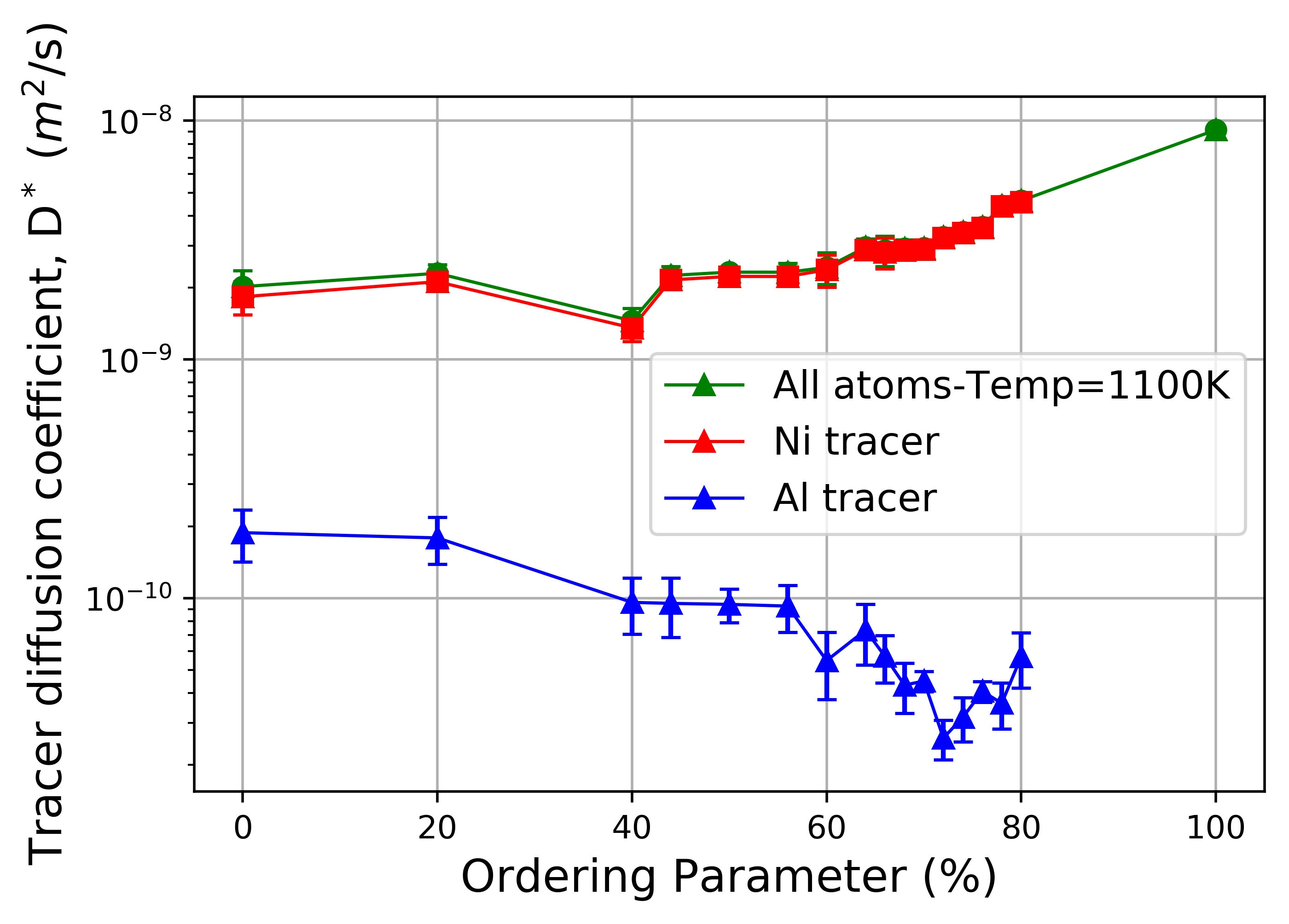}
			\caption{1100K}
		\end{subfigure}
		\caption{Tracer diffusion as a function of initial LROs at six different temperatures.}
		\label{tracer_diff}
	\end{figure*}

\begin{figure*}[bht]
		\centering
		\begin{subfigure}[b]{.45\linewidth}
			\includegraphics[scale=0.5]{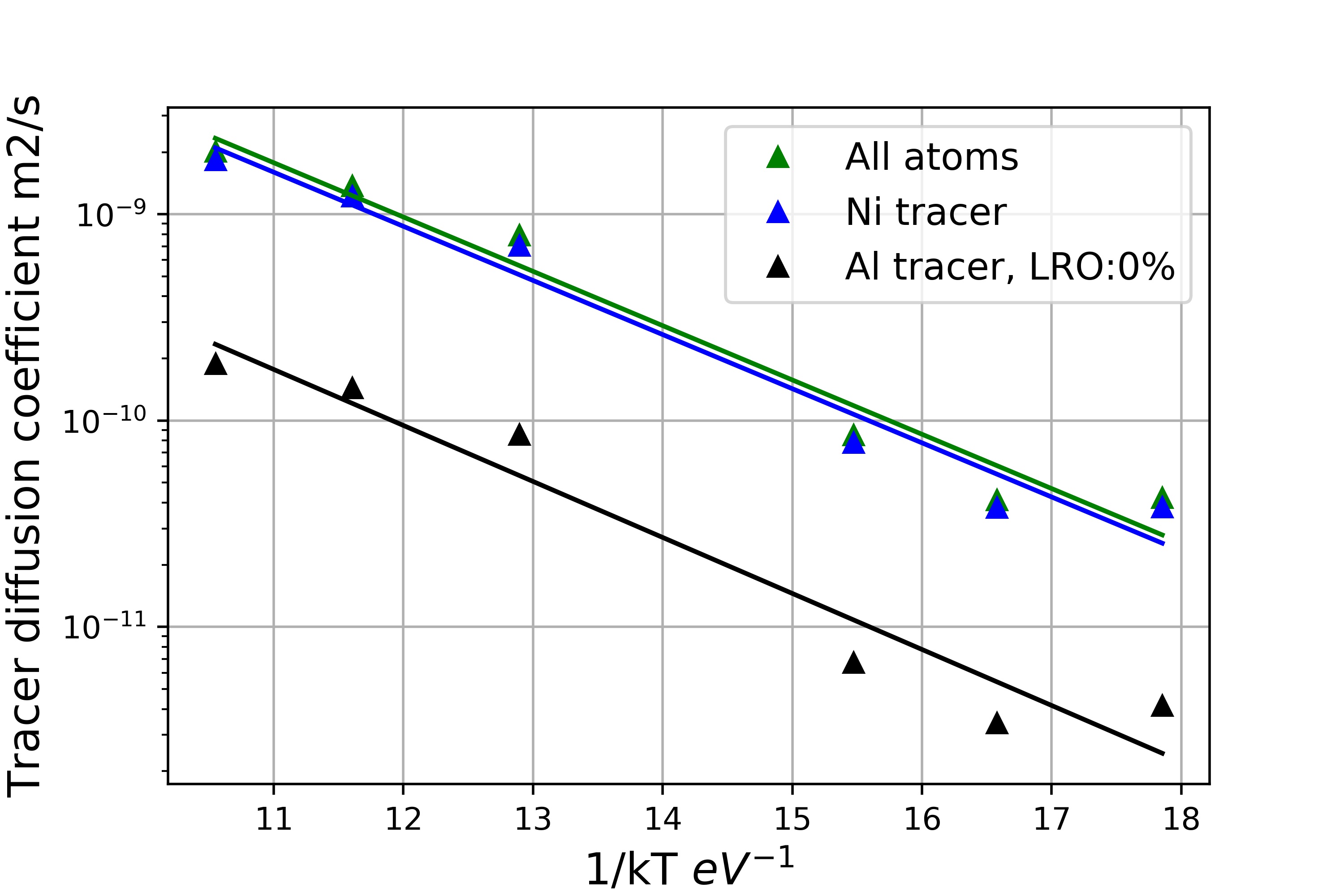}
			\caption{LRO:0\%}
		\end{subfigure}
		\begin{subfigure}[b]{.45\linewidth}
			\includegraphics[scale=0.5]{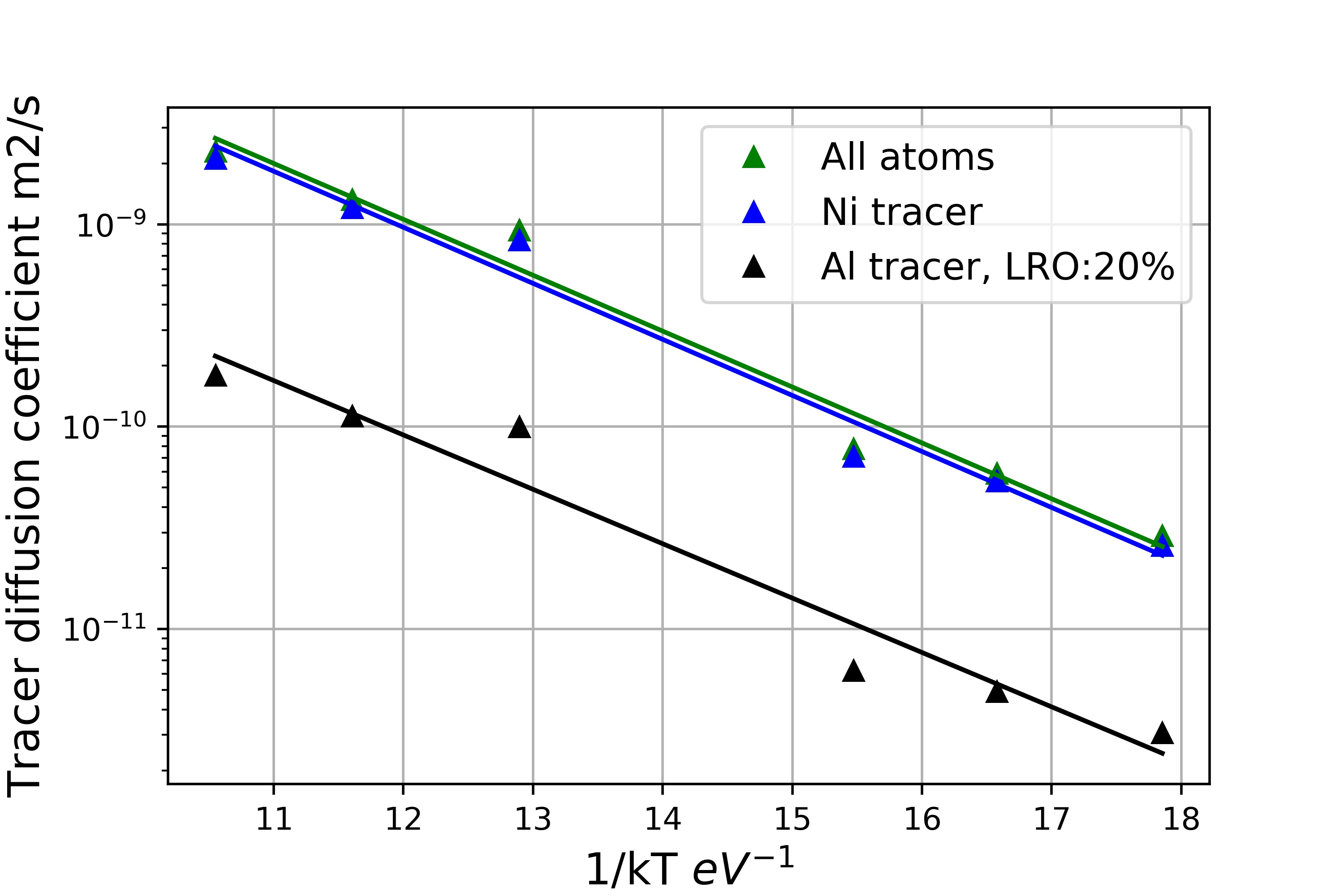}
			\caption{LRO:20\%}
		\end{subfigure}\\
		\begin{subfigure}[b]{.45\linewidth}
			\includegraphics[scale=0.5]{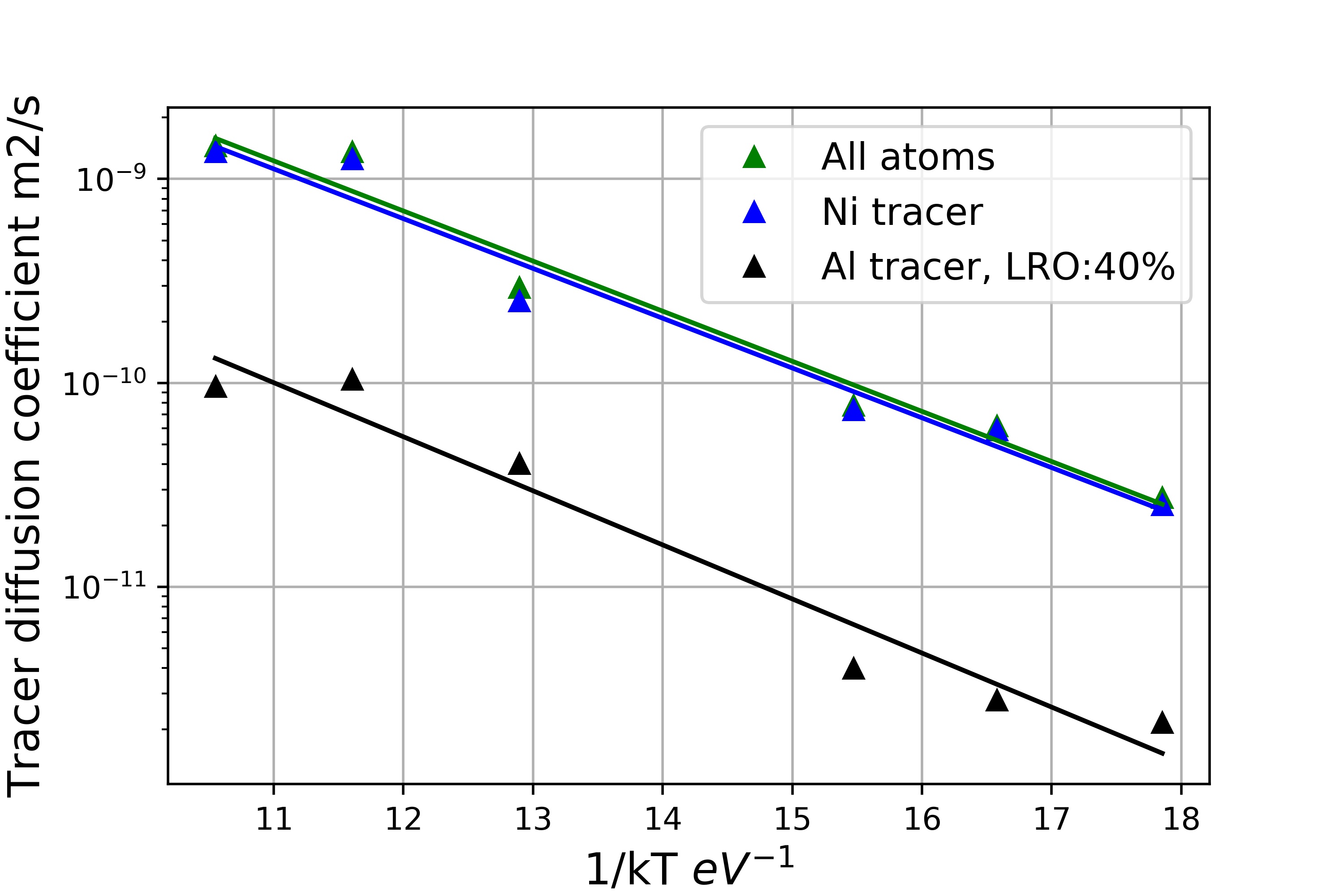}
			\caption{LRO:40\%}
		\end{subfigure}
		\begin{subfigure}[b]{.45\linewidth}
			\includegraphics[scale=0.5]{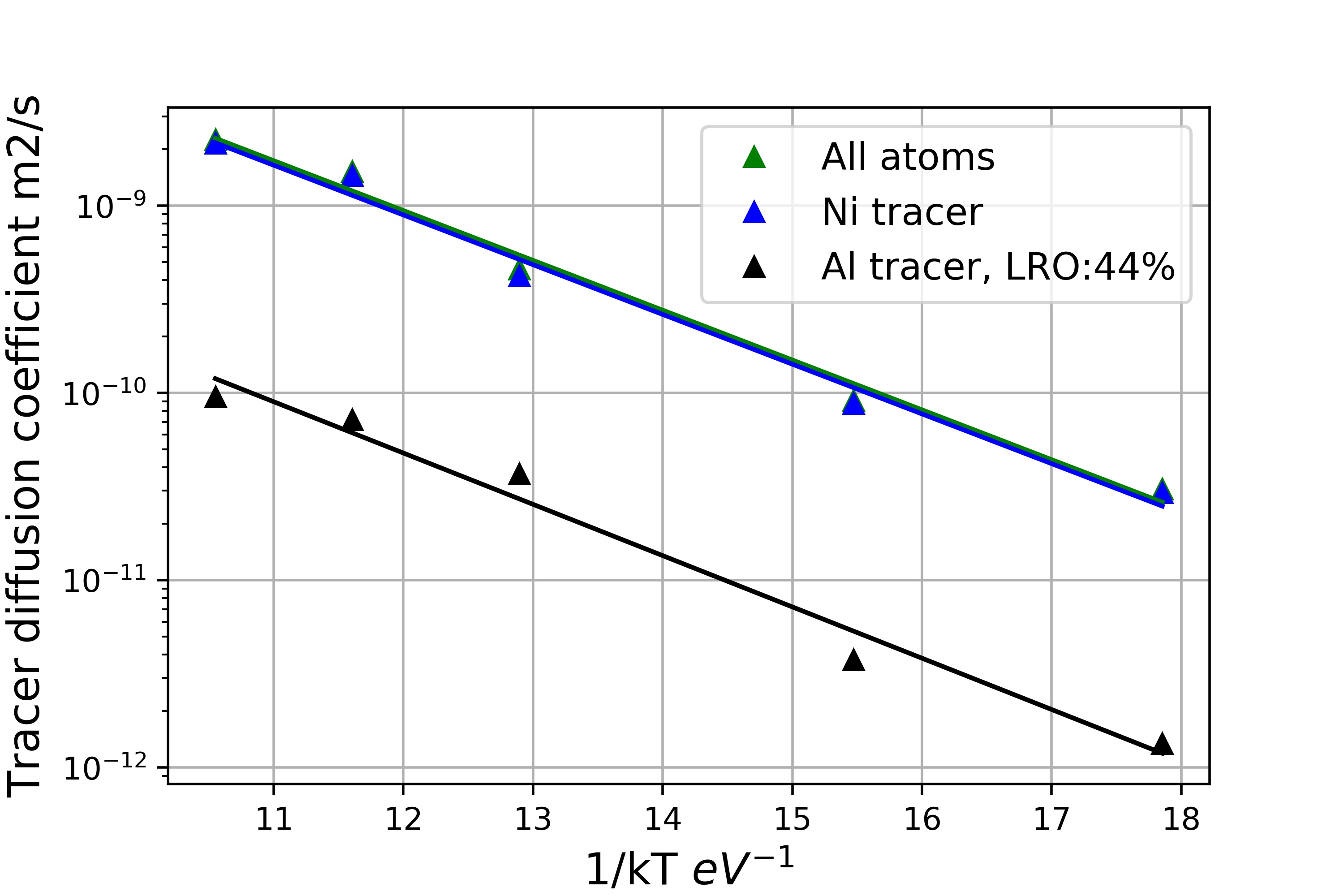}
			\caption{LRO:44\%}
		\end{subfigure}\\
		\begin{subfigure}[b]{.45\linewidth}
			\includegraphics[scale=0.5]{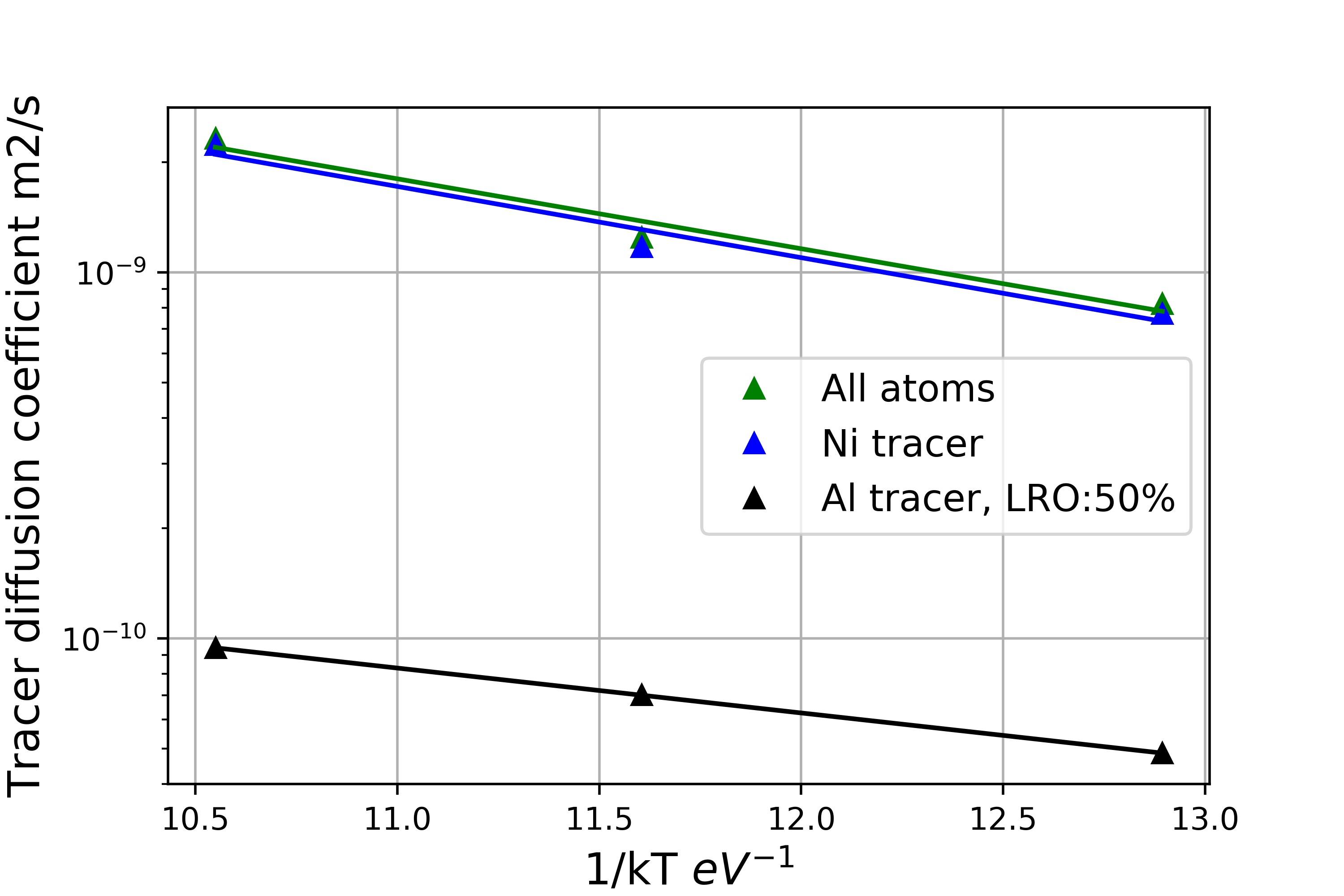}
			\caption{LRO:50\%}
		\end{subfigure}
		\begin{subfigure}[b]{.45\linewidth}
			\includegraphics[scale=0.5]{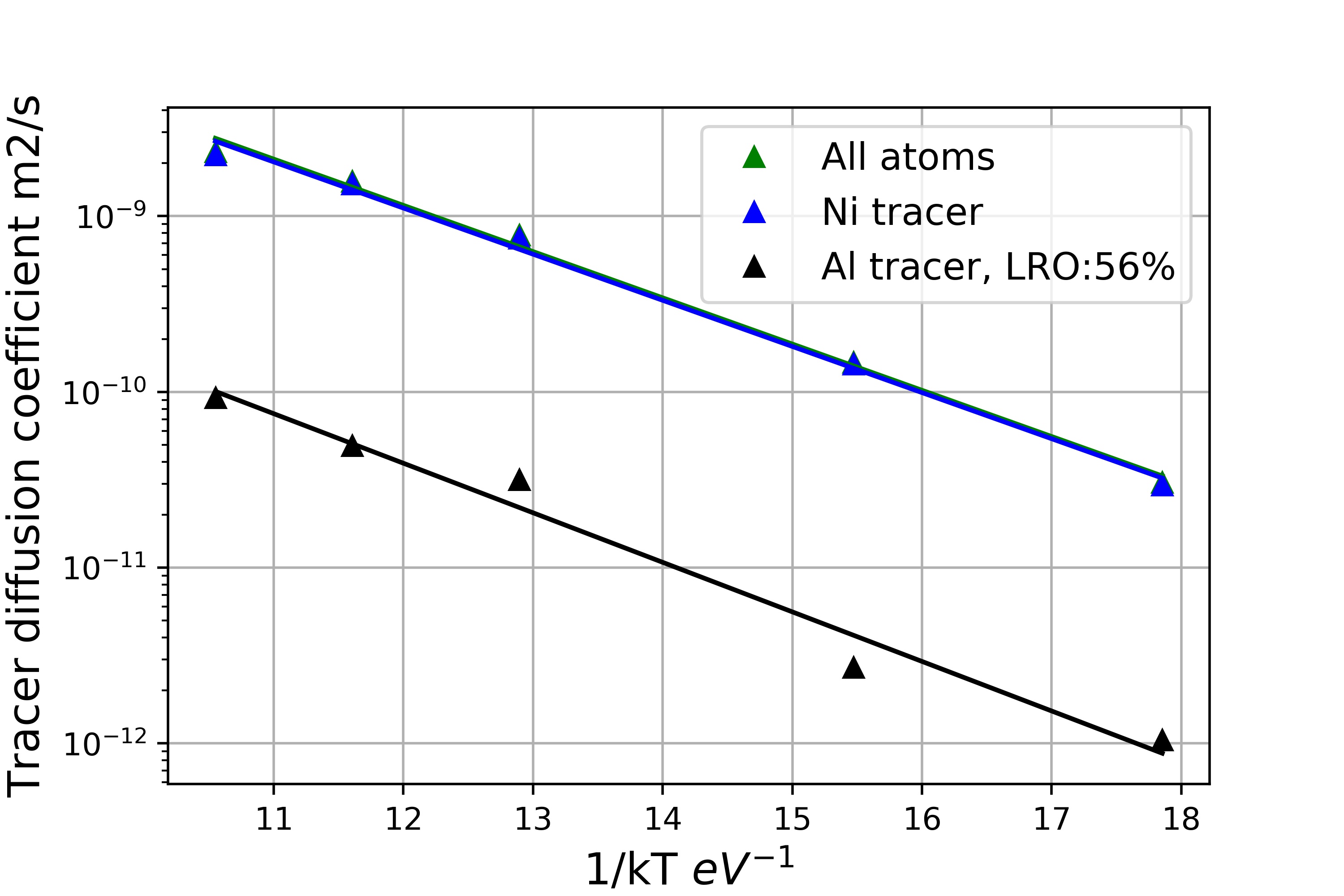}
			\caption{LRO:56\%}
		\end{subfigure}\\
		\begin{subfigure}[b]{.45\linewidth}
			\includegraphics[scale=0.5]{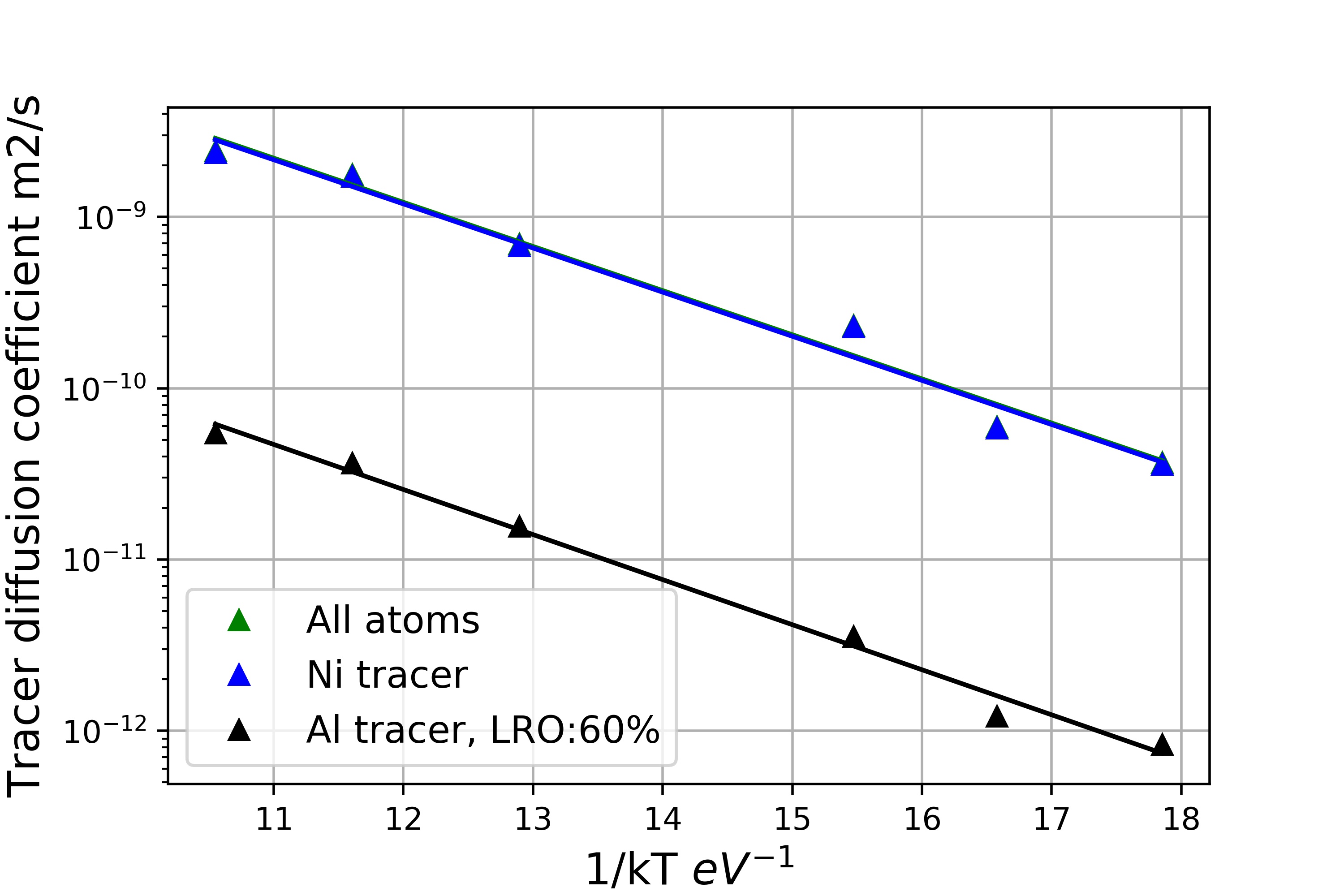}
			\caption{LRO:60\%}
		\end{subfigure}
		\begin{subfigure}[b]{.45\linewidth}
			\includegraphics[scale=0.5]{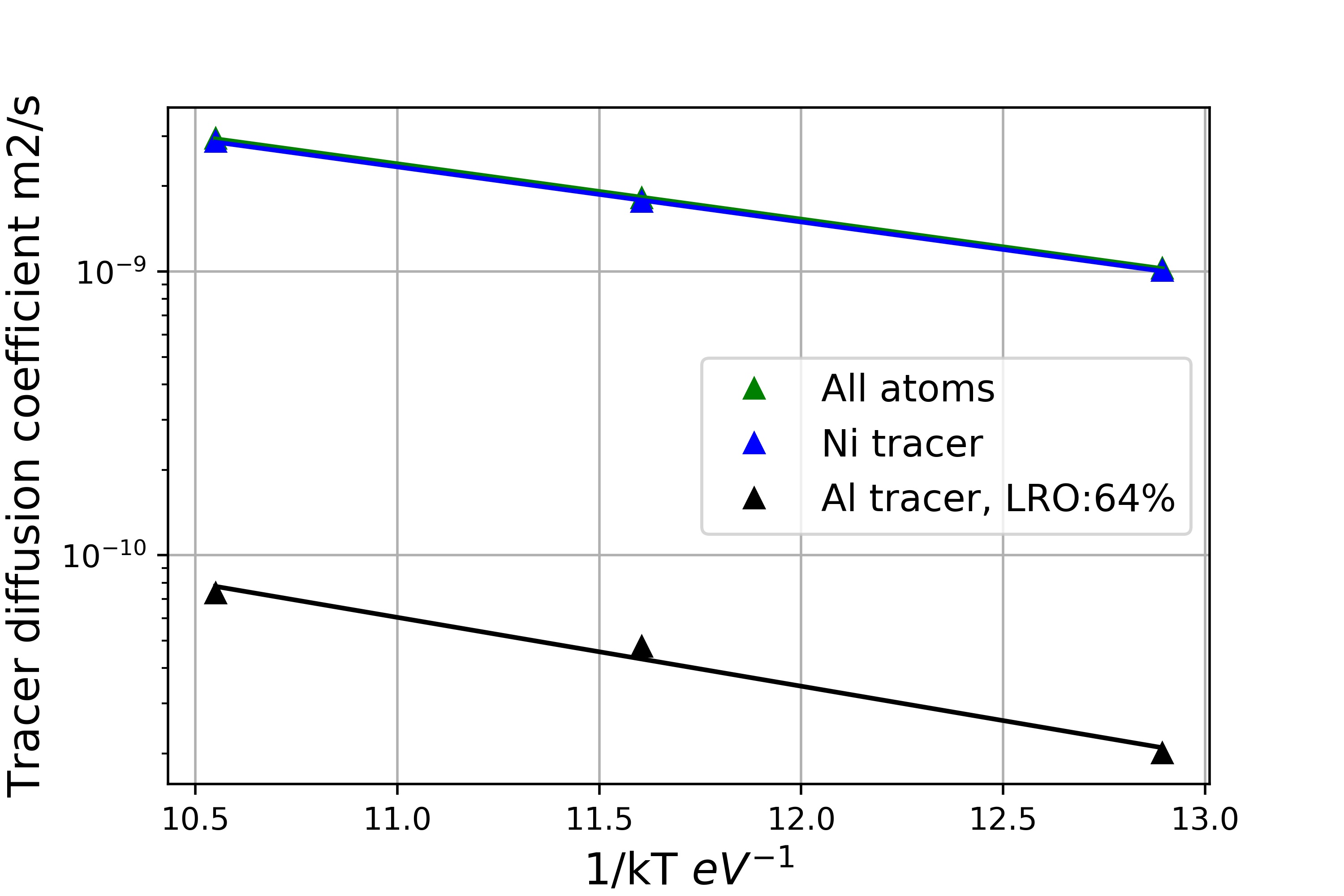}
			\caption{LRO:64\%}
		\end{subfigure}
		\caption{Tracer diffusion coefficient versus inverse of temperature at different initial LROs.}
		\label{Diff}
	\end{figure*}

	Here are the steps to calculate tracer diffusion coefficients and effective activation barrier. Figures \ref{Square1}-\ref{Square5} show the square displacement as a function of time at different temperatures. The slope of the lines is tracer diffusion coefficient as indicated in figure \ref{tracer_diff}. The effective activation barrier, Q$_{eff}$, is calculated using the diffusion coefficient at different temperatures. The Arrhenius plots for temperature dependence of total, $D^*$, and partial, ${D}_{Ni}^*$ and ${D}_{Al}^*$ tracer diffusion coefficients for LROs up to 64\% are shown in figure \ref{Diff}.
	
	\section{SRO model with lower Q$_{eff}$}
	
	A lower migration barrier, Q$_{eff}$, from Zhao's study \cite{zhao2021structural} is used in the model, and it gives quite a similar prediction at 500 K.
	
	\begin{figure*}[!ht]
		\centering
		\begin{subfigure}[b]{.4\linewidth}
			\includegraphics[scale=0.45]{SRO_time.jpg}
			\caption{Q$_{eff}$ from the current study.}
		\end{subfigure}
		\begin{subfigure}[b]{.4\linewidth}
			\includegraphics[scale=0.45]{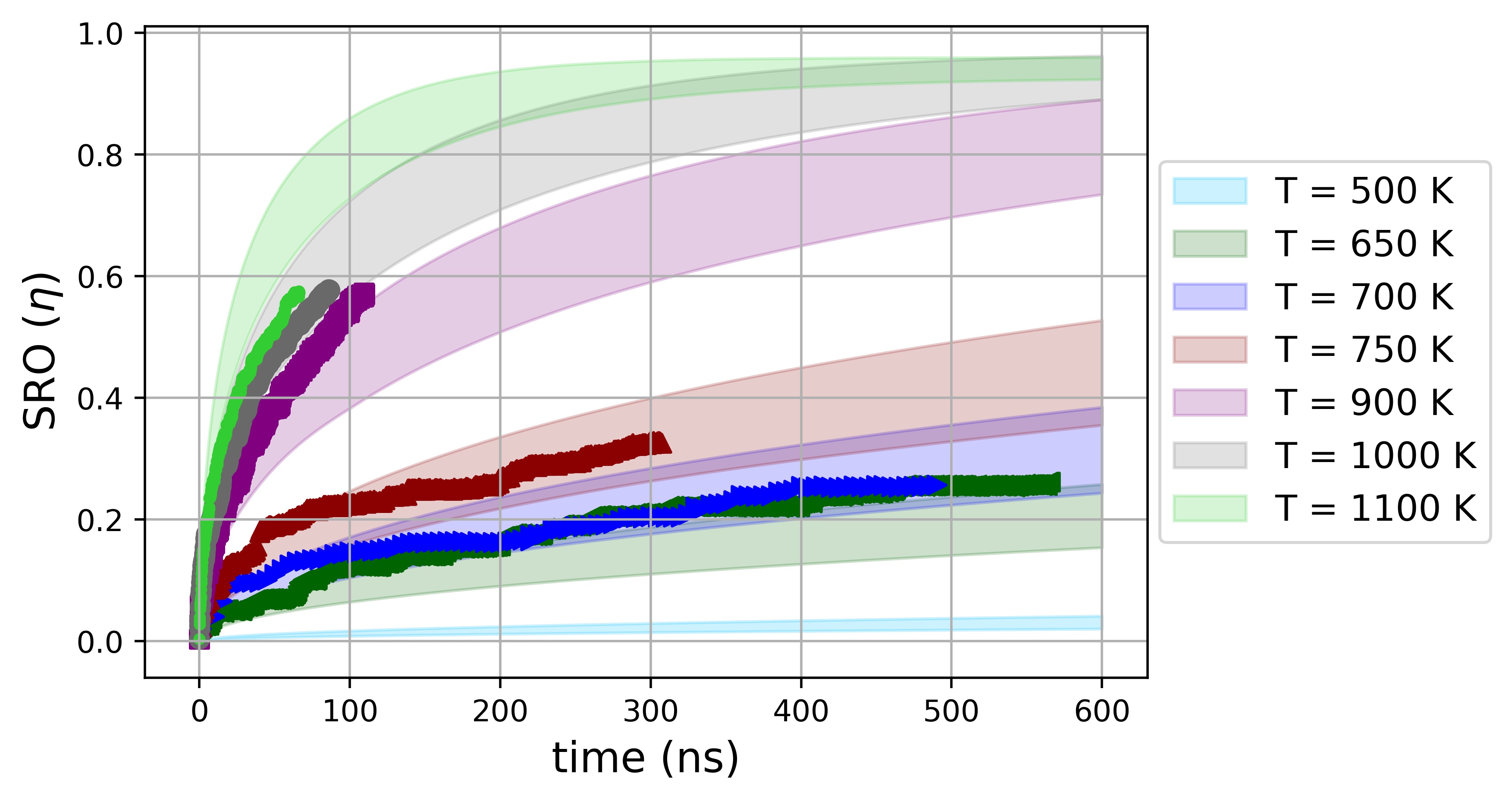}
			\caption{Q$_{eff}$ from Zhao's work \cite{zhao2021structural}.}
		\end{subfigure}
		\caption{Analytical model prediction for two different Q$_{eff}$. The narrow band at the bottom of graph shows the prediction at 500 K.}
		\label{model}
	\end{figure*}
	
	\section{Number of $<$100$>$ self-interstitial dumbbells throughout the simulation}
	
	The number $<$100$>$ self-interstitial dumbbells are counted at three different temperatures to check the configurational state of the SIA throughout the simulations. 2000 dump files at 650 \& 750 K and 10000 dump files at 700 K are used to check the configurational state of the SIA. Figure 22 shows that more than 70-80 \% of the time, the SIA is a $<$100$>$ dumbbell. A self-interstitial dumbbell with an angle less than 15 degrees from $<$100$>$ directions is considered to be a $<$100$>$ dumbbell.

	\begin{figure*}[!ht]
		\centering
		\begin{subfigure}[b]{.4\linewidth}
			\includegraphics[scale=0.45]{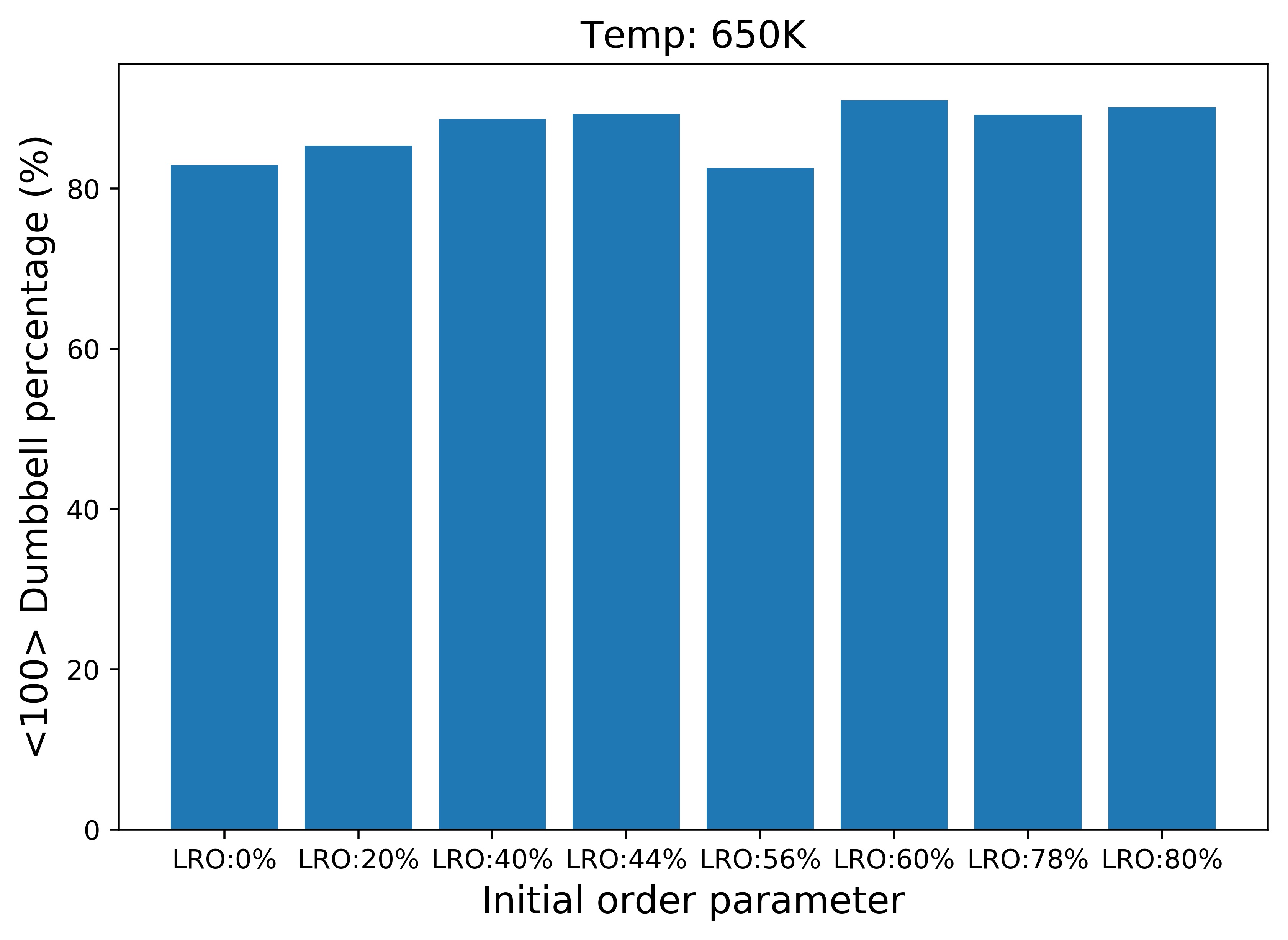}
		\end{subfigure}\\
		\begin{subfigure}[b]{.4\linewidth}
			\includegraphics[scale=0.45]{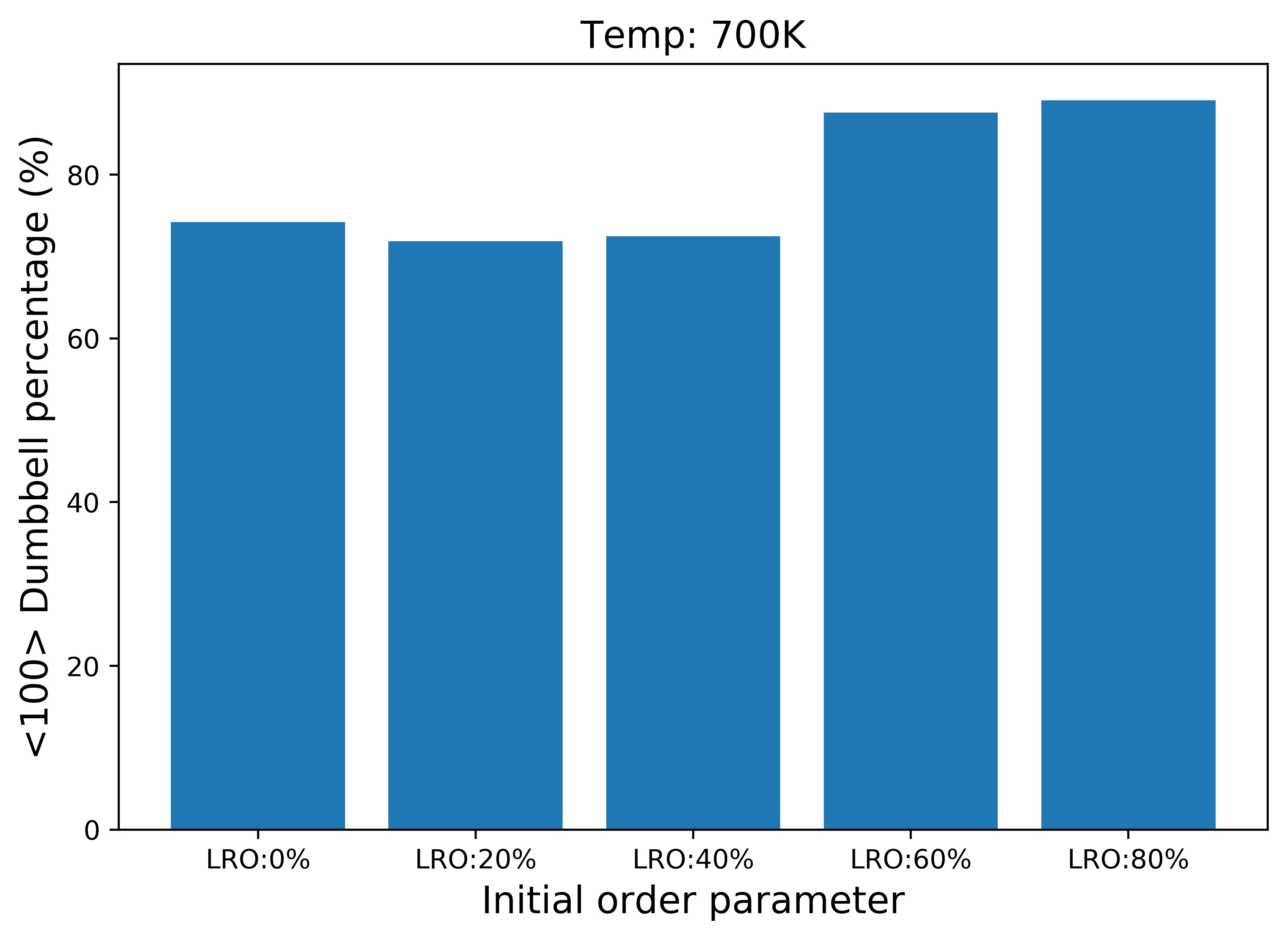}
		\end{subfigure}
		\begin{subfigure}[b]{.4\linewidth}
			\includegraphics[scale=0.45]{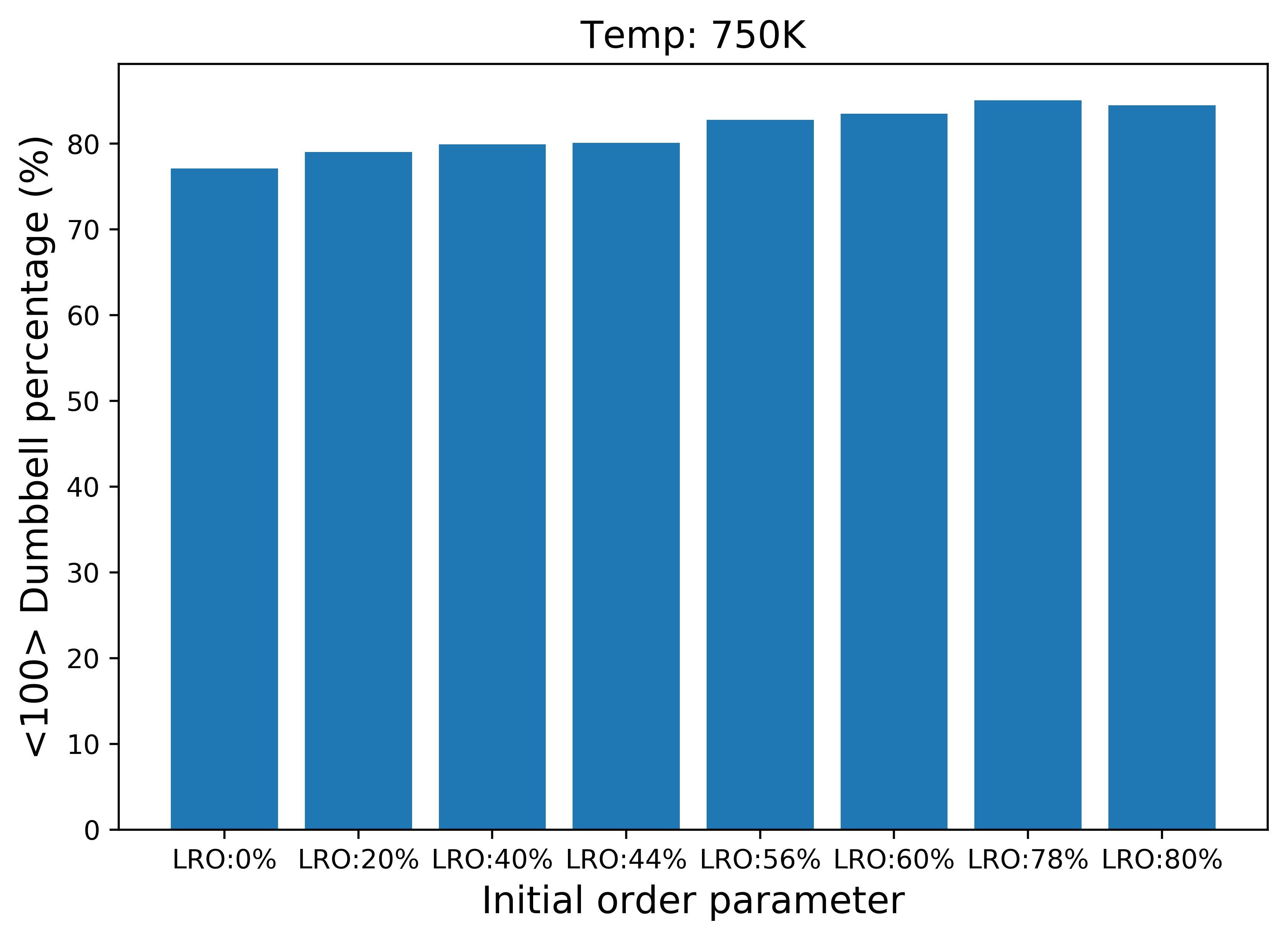}
		\end{subfigure}
		\caption{It is observed that more than 70-80\% of times throughout the simulation the SIA has a $<$100$>$ dumbbell configuration.}
		\label{Dumbbell_config}
	\end{figure*}
	
	\bibliographystyle{unsrt}
	\bibliography{sm}